\documentclass[11pt,letterpaper]{article}
\pdfoutput=1
\usepackage{jcappub}
\usepackage{xcolor}
\usepackage{caption}
\usepackage{ragged2e}
\usepackage{tabularray}
\UseTblrLibrary{booktabs}
\usepackage{ragged2e}
\usepackage{amsmath}
\usepackage{amssymb}
\usepackage{cases}
\usepackage{dsfont}
\usepackage{subfigure}
\usepackage{graphicx}
\usepackage{bm}
\usepackage{color}
\usepackage{booktabs}
\usepackage[mathlines]{lineno}

\newcommand{\g}{g^2}

\newcommand{\Avec}{{\bm A}}

\newcommand{\Hvec}{{\bm H}}

\newcommand{\Xvec}{{\bm X}}
\newcommand{\xvec}{{\bm x}}
\newcommand{\kvec}{{\bm k}}

\newcommand{\yvec}{{\bm y}}
\newcommand{\zvec}{{\bm z}}

\newcommand{\Ocal}{\mathcal{O}}

\newcommand{\Lscr}{\mathscr{L}}

\newcommand{\ba}[1]{\begin{align} #1 \end{align}}

\newcommand{\bsa}[2]{\begin{subequations}\label{#1}\begin{align} #2 \end{align}\end{subequations}}

\newcommand{\Beq}{\begin{equation}\begin{aligned}}
\newcommand{\Eeq}{\end{aligned}\end{equation}}

\usepackage{graphicx, color}
\usepackage{dcolumn}
\usepackage{bm}
\usepackage[mathlines]{lineno}
\usepackage{amsmath}
\usepackage{amssymb}
\usepackage[numbers,sort&compress]{natbib}
\usepackage{bm}
\usepackage{float}
\usepackage{mathrsfs}
\usepackage{lipsum}
\usepackage{hyperref}
\usepackage[normalem]{ulem}
\hypersetup{
    colorlinks=true,       
    linkcolor=rp,          
    citecolor=green2,        
    filecolor=magenta,      
    urlcolor=green2           
}

\usepackage[all]{hypcap} 
\usepackage{cleveref}
\definecolor{rp}{cmyk}{0.2, 1, 0.6, 0}
\definecolor{rp}{cmyk}{0.2, 1, 0.6, 0}
\definecolor{green2}{cmyk}{0.27, 0, 1, 0.52}

\usepackage{float}
\usepackage{cancel}
\usepackage{soul}
\usepackage{cases}
\usepackage{dsfont}
\usepackage{graphicx}
\usepackage{bm}
\usepackage{color}
\usepackage{amsmath}
\usepackage{adjustbox}
\usepackage{amsmath}
\usepackage{amssymb}
\usepackage{mathrsfs}
\usepackage{textcase}
\usepackage{colortbl}
\usepackage{ulem}
\usepackage{mathrsfs}     
\newcommand{\beqs}{\begin{eqnarray}}
\newcommand{\eeqs}{\end{eqnarray}}

\title{\huge Dark Spin-2 Field Solitons as a Source of Electromagnetic Radiation}

\author{Enrico D. Schiappacasse} 
\emailAdd{Enrico.Schiappacasse@uss.cl}

\affiliation{Facultad de Ingenier\'ia, Universidad San Sebasti\'an, Bellavista 7, Santiago 8420524, Chile}

\date{\today}

\abstract{Spin-$s$ light dark boson particles can exhibit wave-like behavior, capable of forming long-lived, coherent, spatially localized structures known as solitons. This work considers the possibility that a light spin-2 particle might be part of or all the dark matter content of the Universe, which could result in a significant fraction of solitons existing today in galactic halos. If these dark matter particles interact with electromagnetism through dimension-6 operators, the solitons may experience parametric resonance of photons triggered by the surrounding electromagnetic field. We explore the feasibility and key characteristics of this electromagnetic radiation, as well as the potential for detection through soliton mergers using ground-based facilities.\\}

\begin{document}

\maketitle

\newpage

\section{Introduction}

One of the most significant mysteries in modern physics is the true nature of dark matter (DM). Despite overwhelming cosmological and astrophysical evidence supporting its existence \cite{Peebles:2013hla, Freese:2017idy, Bertone:2016nfn}, we still do not know the spin, charge, or mass of the hypothetical particle or particles that compose it. 

We understand that dark matter is currently non-relativistic, aggregates under the influence of gravity, and interacts weakly with Standard Model particles. Furthermore, suppose the dark matter particle is bosonic and sufficiently light. In this case, the related field can be described classically as a result of its high occupancy number and the overlap of de Broglie wavelengths. In this framework, dark matter is known as wave dark matter, with a mass much less than one electronvolt, $m \ll\,\text{eV}$\,\cite{Hui:2021tkt}. This type of dark matter is linked to intriguing and novel phenomena that stem from its wavelike characteristics, including interference, coherence, and resonance.

One of the most striking predictions regarding light bosonic dark matter is the formation of long-lived, coherent, spatially localized structures known as solitons. In recent years, these solitons have garnered significant attention from the physics community because of their unique properties, which suggest that they could serve as potential astrophysical laboratories within galaxies. 

The mass-radius relation for solitons is described by the equation \(M_{\text{sol}}R_{\text{sol}} \sim 200\, (m^2_{\text{pl}}/m^2)\) \cite{Jain:2021pnk}, where \(m_{\text{pl}}\) represents the reduced Planck mass. For a particle mass of approximately \(m \sim 10^{-6}\, \text{eV}\),  solitons with a typical mass of about \(M_{\text{sol}} \sim 10^{-9}\,M_{\odot}\) will have  a radius of \(R_{\text{sol}} \sim 200\,\text{km}\). These solitons exhibit an average density given by the expression \(3M_{\text{sol}}/(4\pi R^3_{\text{sol}}) \sim 10^{-8}\,M^4_{\text{sol}} \left(m/m_{\text{pl}}\right)^6\), which is roughly \(O(10^{26})\) times greater than the local dark matter density \cite{Read:2014qva}. Essentially, we are discussing small astrophysical compact objects that contain a high density of dark matter and show significant resistance to destruction from galactic tides. Because of this high density, any weak interactions between dark matter and Standard Model particles may be enhanced within these objects, potentially providing new opportunities for indirect searches for dark matter.

\textcolor{black}{The concept of a massive graviton has attracted considerable interest from both theoretical and phenomenological perspectives since Fierz and Pauli developed the linear theory for a massive spin-2 field\,\cite{Fierz:1939ix}. This linear framework subsequently progressed into a ghost-free nonlinear massive gravity theory and was later expanded to encompass bigravity\,\cite{Hassan:2011zd} and multi-gravity theories\,\cite{Hinterbichler:2012cn}. The spin-2 field has been proposed as a potential dark matter candidate through various realizations. One critical limitation for these realizations is known as the Higuchi bound, which establishes a minimum mass requirement for the spin-2 mass, specifically  $m^2 \geq  2H^2$ (for a De Sitter spacetime)\,\cite{Higuchi:1986py}, where $H$ represents the Hubble parameter. The significance of this bound is contingent upon the period during which the massive graviton is produced\,\cite{Blas:2024kps}; masses below this limit lead to instabilities (Higuchi ghost).} 

\textcolor{black}{For instance, Ref.\,\cite{Aoki:2016zgp} explores the possibility of a heavy graviton being a suitable candidate for dark matter, which is generated during the reheating phase within the ghost-free bigravity framework. They require that the mass of the graviton must be greater than the Hubble expansion rate at the time of its creation to avoid the Higuchi instability. Consequently, the mass range is limited to $10^{-4}\,\text{eV}\lesssim m \lesssim 10^{7}\,\text{eV}$. On the other hand,
Ref.\,\cite{Kolb:2023dzp} addresses the cosmological gravitational production of spin-2 particles in hilltop inflation. In the context of ghost-free bigravity, and by generalizing the Higuchi instability to a Friedmann-Robertson-Walker spacetimes, the authors find that massive graviton gravitational production can account for all of the dark matter in the mass range of  $O(1) \lesssim m/H_{\text{inf}} \lesssim O(10)$ (reheating temperature about $(10^2-10^6)\,\text{GeV}$, with $H_{\text{inf}}$ as the inflationary Hubble parameter).}

\textcolor{black}{In related studies, authors of Refs.\,\cite{Gorji:2023cmz, Gorji:2025tos} examine the production of spin-2 particles during inflation within the framework of a ghost-free effective field theory (EFT) for a light, massive graviton. By considering a strong coupling between the field and the inflaton, they introduce deviations from the exact symmetries of de Sitter spacetime, which allows for the relaxation of the Higuchi bound (this bound arises from these symmetries, as discussed in \,\cite{Kehagias:2017cym, Bordin:2018pca}). In the specific case of a scale-dependent power spectrum for the spin-2 field, which shows a sharp peak at sub-CMB scales, and taking into account constraints from backreaction, Big Bang Nucleosynthesis, and stability, the EFT for a massive graviton can explain dark matter across a wide mass range, specifically from $10^{-21}\,\text{eV}$ to $H_{\text{inf}}$ (see Fig. 3 in \cite{Gorji:2025tos})\footnote{\textcolor{black}{As the spin-2 field spectrum is completely suppressed at CMB scale, there is no isocurvature bound\,\cite{Gorji:2025tos}.}}. Although the model assumes that the massive graviton exists in the dark sector and is decoupled from standard model particles, this assumption can be relaxed. One may consider the possibility of direct interactions between the spin-2 field and other fields during and after inflation\,\cite{Gorji:2023cmz}.}

\textcolor{black}{The effective non-relativistic theory of gravitationally interacting, massive bosonic spin-$s$ fields, which allow for soliton configurations, is performed in Ref.\,\cite{Jain:2021pnk}. In this work, the fields are considered part of or all dark matter. Since the focus is on the non-relativistic limit, the authors present a general action up to quadratic order in the fields, along with leading-order gravitational interactions, explicitly showing the existence of scalar, vector, and tensor solitons}. These solitons correspond to the ground state solutions of the (2$s$ + 1) component Schrödinger-Poisson system, where \( s \) represents the spin of the non-relativistic light bosonic field. The primary distinction between scalar solitons, which consist of spin-0 particles, and those formed by higher spin bosonic particles lies in the nature of the fields involved. For spin-\( s \) fields, there are \( (2s + 1) \) extremally polarized solitons, where all constituent waves of the soliton share the same spin multiplicity. Consequently, these polarized solitons exhibit a richer structure than scalar solitons, while maintaining the same universal mass density profile. This characteristic provides a unique opportunity for specific phenomenology that may not be available for scalar solitons. 

The formation of scalar solitons has been studied in various contexts, including the early Universe \cite{Guth:2014hsa, Brax:2020oye}, virialized dark matter halos and miniclusters \cite{Levkov:2018kau, Eggemeier:2019jsu}, as well as minihalos surrounding primordial black holes \cite{Hertzberg:2020hsz, Yin:2024xov}. For instance, solitons composed of axions can nucleate within typical axion miniclusters at a redshift of approximately \( O(10^2) \), with a nucleation time on the order of \( O(10^7-10^8) \) years \cite{Eggemeier:2019jsu}. Vector soliton formation has also been considered through the collapse of Hubble-scale inhomogeneities at the time of radiation-matter equality \cite{Gorghetto:2022sue} and through the merging of halos or solitons \cite{Amin:2022pzv}.

\textcolor{black}{In the context of the work presented in Ref.\,\cite{Levkov:2018kau}, which demonstrates Bose-Einstein condensation in the kinetic regime of scalar solitons within virialized dark matter halos or miniclusters through universal gravitational interactions, Ref. \cite{Jain:2023ojg} expands on this research to explore the gravitational condensation and formation of solitons made up of spin-$s$ dark matter particles. Specifically, the authors focus on spin values \(s=0,1,2\), corresponding to scalar, vector, and tensor solitons, respectively. 
Levkov, Panin, and Tkachev in Ref.\,\cite{Levkov:2018kau} indicate that the formation of scalar solitons occurs through kinetic relaxation over a typical timescale given by $\tau_\mathrm{gr} \sim (m_\mathrm{pl}^4 m^3 v^6)/[\rho_\mathrm{halo}^2 \log(mvR_\mathrm{halo})]$. Here, \(m_\mathrm{pl}\) represents the reduced Planck mass, while \(v\) and \(\rho_\mathrm{halo}\) pertain to the characteristic velocities and mass density of dark matter in the halo, respectively, with \(R_\mathrm{halo}\) denoting the size of the halo. 
In their study, the authors of Ref.\,\cite{Jain:2023ojg} investigate a (2$s$ + 1) component Schrödinger-Poisson system, conducting approximately 100 numerical simulations. Their findings indicate that gravitational Bose-Einstein condensation serves as a universal mechanism for forming scalar, vector, and tensor solitons, operating under the same underlying principles and exhibiting similar nucleation timescales. Specifically, solitons consisting of light spin-1 or spin-2 particles are found to nucleate within virialized halos in a period roughly equal to $O(1)\tau_\mathrm{gr}$.}

For spin-0 particles, solitons are associated with explosive events involving relativistic particles resulting from gravitational collapse \cite{Eby:2016cnq, Di:2024jia, Levkov:2016rkk}, dipole radiation \cite{Amin:2021tnq}, and radio emission through the parametric resonance of photons \cite{Hertzberg:2018zte, Hertzberg:2020dbk, Levkov:2020txo, Chung-Jukko:2023cow}. In the case of vector solitons, which consist of spin-1 particles, parametric resonance of photons is also present, as discussed in Ref.\,\cite{Amin:2023imi}. There have been no attempts to explore the phenomenology associated with tensor solitons, similar to those conducted for scalar and vector solitons. In Ref.\,\cite{Jain:2021pnk}, the authors briefly mention possible dimension-6 interactions between spin-2 particles and electromagnetism, pointing out that these non-gravitational interactions may aid in the eventual detection of tensor solitons. 

In this work, we explore the phenomenon of parametric resonance in solitons, focusing specifically on the scenario where a light spin-2 particle constitutes all or part of dark matter. Our main objective is to analyze the key characteristics of this phenomenon, discuss the feasibility of potential detection, and compare our findings with established results related to scalar and vector solitons.

To achieve this, we consider a direct coupling between the spin-2 particles and electromagnetism, utilizing a limited set of dimension-6 operators within the effective field theory framework \textcolor{black}{and following the suggested operators given in Ref.\,\cite{Jain:2021pnk}}. Although we do not attempt to address all possible operators that maintain electromagnetic gauge invariance,  our selected operators are sufficient to capture the essential conditions for resonance and its main features, such as frequency, polarization, and spatial radiation patterns. A more comprehensive analysis will be performed for us in future work.

\textcolor{black}{As we will see, the resonance phenomenon is associated with a photon production, which peaks at a frequency given by the mass of the spin-2 particle. Considering possible radio signal detection on Earth, we focus on the spin-2 particle mass range \(10^{-7} \,\text{eV}\lesssim m \lesssim 10^{-3}\,\text{eV}\). As explained before, this dark matter range for the particle mass can be obtained, for instance, by considering spin-2 particle production during inflation via a strong coupling between the field and the inflaton within the framework of a ghost-free effective field theory\,\cite{Gorji:2025tos}. On the other hand, we remark that tensor solitons are expected to nucleate in dark halos via gravitational condensation in the kinetic regime as shown in Ref.\,\cite{Jain:2023ojg}.}

We use natural units throughout the paper by setting $\hbar=c=1$.

\section{Dark spin-2 field and electromagnetism interactions}
\label{Sec:interactions}

\textcolor{black}{The interaction between a massive dark tensor field \( H_{\mu\nu}(x) \) and photons \( A_{\mu}(x) \) provides an opportunity to study electromagnetic radiation emitted by dark tensor solitons. Although we operate within the effective field theory (EFT) framework for interactions between spin-2 particles and electromagnetism in the graviton mass range of interest,  it is important to note that this range can be obtained, for instance, from the production of massive gravitons during inflation in a ghost-free effective field theory, as discussed in Ref.\,\cite{Gorji:2025tos}. In this work, the authors assume that the spin-2 field is the only other field present, besides the inflaton, during the inflationary period, and that it exists within the dark sector of the universe, meaning that it is decoupled from standard model particles. However, these assumptions can be relaxed; one may consider direct interactions between the spin-2 field and other fields during and after inflation. Such interactions would introduce corrections to the parameters of the theory, which could be adjusted to maintain stability and prevent ghost-like properties\,\footnote{\textcolor{black}{We would like to thank Mohammad Ali Gorji for providing clarification on this point during a private communication.}}.}

\subsection{Effective field theory interactions with electromagnetism}

Interaction between a massive dark tensor field $H_{\mu\nu}(x)$ and photons $A_{\mu}(x)$ opens a window for the study of electromagnetic radiation from dark tensor solitons. In the effective field theory (EFT) framework, the action for photons and the interacting term between spin-2 particles and them is the following:
\begin{equation}
\mathcal{S}_{A_{\mu},A_{\mu}-H_{\mu\nu},\textcolor{black}{\bar g_{\mu\nu}}} = \int\text{d}^{4}x\sqrt{-\textcolor{black}{\bar g}}\left[ \frac{1}{2}m^2_{\text{pl}}R -\frac{1}{4}F_{\mu\nu}F^{\mu\nu} +\Lscr_{\text{int}}  \right ]\,,
\label{eq:Action}
\end{equation}
where  $F_{\mu\nu} = \partial_{\mu}A_{\nu}-\partial_{\nu}A_{\mu}$ is the usual electromagnetic field stress tensor, $R$ is the Ricci scalar, the diagonal inverse metric takes a signature $(-,+,+,+)$, \textcolor{black}{$\bar g \equiv |\bar g_{\mu\nu}|$,}  and the Levi-Civita symbol is defined by $\epsilon^{0123}=1$.  The spin-2 boson-photon interacting Lagrangian density is defined by $\Lscr_\mathrm{int} = g^2 \Ocal_i$, with $\Ocal_i$ a defined dimension-6 operator,  and $g^2$ the coupling\,\footnote{\textcolor{black}{We would like to highlight that dimension-5 operators of the form \(F_{\mu\nu}F_{\alpha\beta}H_{\sigma\theta}\bar{g}_{\gamma\epsilon}\)\,\cite{deRham:2010ik} present an intriguing possibility that we will explore in future work. In this study, we have concentrated on dimension-6 operators, following the approaches proposed in Ref.\,\cite{Jain:2021pnk}, and we aim to compare our results with those already obtained in vector solitons, where dimension-6 operators are considered\,\cite{Amin:2023imi}.}}. 

To construct operators \(\mathcal{O}_{i}\), we require electromagnetic gauge invariance. There are no dimension-6 operators involving a single factor of the tensor field \(H_{\mu\nu}\) because they carry an odd number of Lorentz indices, making a full contraction impossible. Therefore, we may consider three families of operators: 
 \( F_{\mu\nu} F_{\alpha\beta} H_{\sigma\theta} H_{\gamma\epsilon}\)\,,  
\( F_{\mu\nu}  H_{\alpha\beta} H_{\sigma\theta} H_{\gamma\epsilon} H_{\delta\xi}\)\,, and\, 
\( F_{\mu\nu} \partial_{\alpha} \partial_{\beta} H_{\sigma\theta} H_{\gamma\epsilon}\).
However, the second and third families include only one factor of the electromagnetic field \(A_{\mu}\), which means these operators would act as sources for \(A_{\mu}\). Source operators have been previously studied in the context of scalar solitons\,\cite{Amin:2021tnq}, demonstrating that the emitted radiation power is exponentially suppressed by a factor of $mR_{\text{sol}} \gg 1$. We expect that this characteristic will also hold for the tensor soliton case. 

Focusing on the first family, the available operators for a spin-0 field 
$\phi(x)$ are given by: $\mathcal{\bar{O}}_1 = -(1/4)\phi F_{\mu\nu} \tilde{F}^{\mu\nu} $ and $ \mathcal{\bar{O}}_2 = -(1/4)\phi F_{\mu\nu} F^{\mu\nu}$. This fact makes it particularly interesting to study the spin-2 field operators where two spin-2 fields are fully contracted with each other. For the construction of the non-relativistic tensor soliton field in Eqs.\,(\ref{Eq:H2}-\ref{Eq:H0}), we may anticipate unpolarized radiation output for resonance. 
Thus, we should also consider operators in which the Lorentz indices of the spin-2 field are mixed with those of the electromagnetic tensor, following Einstein summation. This feature could lead to a potential polarized radiation output in the context of resonance.  
The reduced pool of dimension-6 operators that we have considered includes the following:
\begin{align}
\Ocal_1 = -\frac{1}{2}F_{\mu\nu}&\tilde{F}^{\mu\nu}H_{\alpha\beta}H^{\alpha\beta}\,,\hspace{0.2cm}
\Ocal_2 = -\frac{1}{2}F_{\mu\nu}F^{\mu\nu}H_{\alpha\beta}H^{\alpha\beta}\,,\hspace{0.2cm}
\Ocal_3 = F_{\mu\nu}F^{\alpha\nu}H_{\alpha\beta}H^{\beta\mu}\,.
\label{Eq:Ops}
\end{align}
Although there are additional operators in the family \( F_{\mu\nu} F_{\alpha\beta} H_{\sigma\theta} H_{\gamma\epsilon} \),  the operators that we have selected are sufficient to capture the essential features of the resonance. A more comprehensive analysis will be conducted in future work. 

In the non-relativistic regime, we are interested in a cold spin-2 field such that the characteristic momenta is much less than the mass, e.g. $k \ll m$. Since $\vert H_{00}\vert =O(m^{-1}) \vert \partial^iH_{i0} \vert = O(m^{-2})\vert \partial^i\partial^jH_{ij} \vert$ with $\vert\partial^iH_{i0}\vert =O(k) \vert H_{i0} \vert$, 
we
have the inequalities\,\cite{Aoki:2017ixz}
\begin{equation}
\vert H_{00} \vert \ll \vert H_{0i} \vert \ll \vert H_{ij} \vert \,,
\end{equation}
and we can safely set $H_{00}=H_{0i}=0$. \textcolor{black}{Within the non-relativistic approximation, where $\bar\mu/m \ll 1$, we systematically drop gradients of the dark tensor field when compared against its time derivative, since $|\partial^i H_{ij} | \sim k|H_{ij}| \ll m |H_{ij}| \sim |\dot H_{ij}|$. This approximation is consistent with the assumption that the tensor field slowly varies in space.}

For the validity of the EFT, we require the condition
\begin{equation}
g^2\bar{H}^2 \ll 1\,,
\label{eq:EFTcond}
\end{equation}
where $\bar{H}$ is the nonzero vacuum expectation value for the dark tensor field and refers to its typical amplitude. \textcolor{black}{Such a condition enables us to disregard the influence of dimension-8 (and higher-order) operators. Additionally, it prevents non-negligible modifications of lower-order operators compared to those in  Eq.\ (\ref{Eq:Ops}). For example, the operator $\mathcal{O}_2$ contributes to the electromagnetic kinetic term in the action given by Eq.\,(\ref{eq:Action}) through the finite $\bar{H}$.}

\subsection{Dark tensor polarized solitons}
\label{Darktensorpolarizedsolitons}

Here, we briefly discuss the main physics and features of tensor solitons. In the non-relativistic regime, a massive spin-2 field can aggregate under the influence of gravity, leading to the formation of polarized tensor solitons. The behavior of this non-relativistic field is described by a multicomponent  Schr$\ddot{\text{o}}$dinger-Poisson system with \((2s + 1)\) components, where \(s = 2\), which are associated with the five physical degrees of freedom of the tensor massive field. The Schr$\ddot{\text{o}}$dinger-Poisson system supports ground state solutions at a fixed number of particles corresponding to long-lived coherent spatially localized configurations with spherically symmetric profiles\,\cite{Jain:2021pnk}. 

There exist \(2(s + 1) = 5\) different and $\textit{extremally}$ polarized tensor solitons that are degenerate in energy (in the absence of self-interactions), with all waves forming the structure possessing the same spin multiplicity. The total intrinsic spin of these solitons can reach significant values, even on an astrophysical scale, along a given direction \(\hat{n}\). This depends on the spin multiplicity \(\lambda = \left\{ -2, -1, 0, 1, 2 \right\}\) and the total number of particles \(N_{\text{sol}}\), and is expressed as $S_{\text{sol}} = \lambda N_{\text{sol}} \hat{n}$\,\cite{Jain:2021pnk}.

In the framework of the effective non-relativistic theory, there are several symmetries that lead to conserved quantities: the total soliton energy \(E_{\text{sol}}\), the total number of particles \(N_{\text{sol}}\), the rest mass \(M_{\text{sol}} = m N_{\text{sol}}\), the total spin angular momentum \(S_{\text{sol}}\), and the total orbital angular momentum \(J_{\text{sol}}\)\,\cite{Jain:2021pnk}. 

\textcolor{black}{The most general case involves a soliton that is formed by a linear combination of various \textit{extremally} polarized solitons. In the non-relativistic regime, we represent the physical degrees of freedom of the real tensor field \(\Hvec(t, \xvec)\), which has spatial indices, as follows\,\cite{Jain:2021pnk}:}

\begin{equation}
\textcolor{black}{
    \Hvec(t,\xvec)  = \frac{1}{2} \sum_{\lambda} \Bigl[c_{(\lambda)} \, {\bm{\epsilon}}^{(\lambda)}_{\hat{n}}\,\tilde H^{(\lambda)}(t,\xvec) \, e^{-i m t} \,  + \mathrm{h.c.} \Bigr]
    \label{Hdefined1}
     \;,}
\end{equation}
\textcolor{black}{where $\tilde{H}^{(\lambda)}(t,\xvec)$ is a slowly varying complex function in space, but also in time in comparison to the highly oscillatory term $e^{-imt}$, with $m \ll \text{eV}$ as the spin-2 boson light mass. Here h.c. indicates hermitian conjugate, and $c_{(\lambda)}$ are real coefficients such that  $\sum_{\lambda} c_{(\lambda)}^2=1 $. The symbol $\lambda$ encompasses the 5 polarization modes, while ${\bm{\epsilon}}^{(\lambda)}_{\hat{n}}$ are the 5 polarization unit vectors at a given direction ${\hat{n}}$. Since we are interested in ground state configurations for a fixed number of particles, we express $\tilde{H}^{(\lambda)}(t,\xvec)$ as spherically symmetric, stationary solutions of the form $H(r)e^{i (\bar\mu - \varphi_{\lambda}/t) t}$, where $\varphi_{\lambda}$ is a phase, and $H(r)$ is the soliton profile that depends solely on the radius. It is important to note that 
$\bar\mu$ is constant and can be thought as the chemical potential\,\cite{Jain:2021pnk, Schiappacasse:2017ham}. By making this substitution in Eq.\,(\ref{Hdefined1}), we have}

\begin{equation}
\textcolor{black}{
    \Hvec(t,\xvec)   
    = \frac{1}{2} \sum_{\lambda} \Bigl[c_{(\lambda)} \, {\bm{\epsilon}}^{(\lambda)}_{\hat{n}}\,H(r) \, e^{-i (m-\textcolor{black}{\bar\mu} + \varphi_{\lambda}/t) t} \,  + \mathrm{h.c.} \Bigr] 
    \label{Hdefined}
     \;.}
\end{equation}
\textcolor{black}{Within the non-relativistic approximation, the soliton oscillates with an angular frequency $\omega$ that is approximately equal to the mass of the tensor field, namely $\omega \approx m$. The chemical potential provides a minor correction to this frequency, expressed as $\omega = m - \bar\mu$, where $\bar\mu/m \ll 1$.}

 Without loss of generality, we take $\hat{n}=\hat{z}$ and consider the orthonormal set of tensors 
\begin{equation}
    {\bm{\epsilon}}^{(\pm 2)} = \frac{1}{2} 
    \begin{bmatrix}
    1&\pm i & 0\\
    \pm i& -1 & 0\\
    0 & 0 & 0
    \end{bmatrix}\,,
    \hspace{0.2cm}
     {\bm{\epsilon}}^{(\pm 1)} = 
   \frac{1}{2} \begin{bmatrix}
    0&0&1\\
    0&0&\pm i\\
    1&\pm i & 0
    \end{bmatrix}\,,
        \hspace{0.2cm}
     {\bm{\epsilon}}^{(0)} = 
   \frac{1}{\sqrt{6}} 
   \begin{bmatrix}
    -1&0&0\\
    0&-1&0\\
    0&0 & 2
    \end{bmatrix}\,.\label{eq:PolMatrix}
\end{equation}
Using the above equations, the five extremally polarized solitons with spin density along the $z$ axis read as
\begin{equation}
\Hvec^{(\pm 2)}(t,\bm x)= \frac{H(r)}{2}  \begin{bmatrix}
      \text{cos}(\omega t)&\pm\text{sin}(\omega t)&0\\
          \pm\text{sin}(\omega t)&-\text{cos}(\omega t)&0\\
          0&0&0
         \end{bmatrix}\,,
           \label{Eq:H2}
         \end{equation}
         \begin{equation}
\Hvec^{(\pm 1)}(t,\bm x)= \frac{H(r)}{2}  \begin{bmatrix}
      0&0&\text{cos}(\omega t)\\
          0&0&\pm \text{sin}(\omega t)\\
          \text{cos}(\omega t)&\pm\text{sin}(\omega t)&0
         \end{bmatrix}\,, 
           \label{Eq:H1}
          \end{equation}
         \begin{equation}
\Hvec^{(0)}(t,\bm x)= \frac{H(r)}{\sqrt{6}}   \begin{bmatrix}
      -\text{cos}(\omega t)&0&0\\
          0&-\text{cos}(\omega t)&0\\
          0&0&2\text{cos}(\omega t)
         \end{bmatrix}\,.  
         \label{Eq:H0}
\end{equation}
The radial field profile $H(r)$ and the non-dynamical Newtonian potential $\Phi(r)$ solve the time-independent multicomponent Schr$\ddot{\text{o}}$dinger-Poisson system of equations. For each of the five polarization modes, a family of solutions is set by the chemical potential $\textcolor{black}{\bar\mu}$, which defines the tensor soliton amplitude. \textcolor{black}{Numerical solutions are  well
characterized by the following fitting function} \cite{Schive:2014dra, Amin:2022pzv, Amin:2023imi}:

\begin{equation}
H(r) =  \frac{2.04\, m_{\text{pl}} (\textcolor{black}{\bar\mu}/m)}{(1+0.077 (\sqrt{\textcolor{black}{\bar\mu} m}r)^2)^4}\,, \label{eq:Hprofile}   
\end{equation}
\textcolor{black}{which refers to the “universal” radial soliton profile}\,\footnote{\textcolor{black}{The radial profile mentioned above has notable characteristics worth discussing. While \( H'(0) = 0 \) ensures that the Laplacian term in the Schrödinger-Poisson system is finite at the origin, the condition \( H'(r \rightarrow \infty) = 0 \) allows us to define configurations as localized objects, even though they do not have a distinct hard surface \,\cite{Jain:2021pnk, Schiappacasse:2017ham}.}}. Here 
the soliton central amplitude reads as $\bar{H}\equiv 2.04\,m_{\text{pl}} (\textcolor{black}{\bar\mu}/m)$. The ``universal"  total soliton mass, energy, and radius (full width at half maximum) read as~\label{Jain:2021pnk} 
\begin{align}
M_{\text{sol}} &\approx 62.3\,\frac{m_{\text{pl}}^2}{m} \left( \frac{\textcolor{black}{\bar\mu}}{m} \right)^{1/2} \sim 10^{-9}\,M_{\odot} \left( \frac{10^{-6}\,\text{eV}}{m} \right)  \left( \frac{\textcolor{black}{\bar\mu}/m}{10^{-11}} \right)^{1/2}\,,\label{eq:Msol} \\
E_{\text{sol}} &\approx -20.8 \frac{m_{\text{pl}}^2}{m} \left( \frac{\textcolor{black}{\bar\mu}}{m} \right)^{3/2}\,\sim -3 \times 10^{45} \,\text{eV} \left( \frac{10^{-6}\,\text{eV}}{m} \right) \left( \frac{\textcolor{black}{\bar\mu}/m}{10^{-11}} \right)^{3/2}\,,\label{eq:Esol} \\
R_{\text{sol}} &\approx 3.16 \frac{1}{m} \left( \frac{\textcolor{black}{\bar\mu}}{m} \right)^{-1/2}\,\sim 200\,\text{km}\, \left( \frac{10^{-6}\,\text{eV}}{m} \right) \left( \frac{\textcolor{black}{\bar\mu}/m}{10^{-11}} \right)^{-1/2}\,.\label{eq:Rsol}
\end{align}
As shown above, for a particular spin-2 particle mass, particular radial profiles and soliton mass, energy, and radius are obtained by setting a value for the $(\textcolor{black}{\bar\mu}/m)$ ratio, which needs to satisfy the condition $\textcolor{black}{\bar\mu}/m \ll 1$ imposed by the non-relativistic regime as explained before. 
\textcolor{black}{As the chemical potential decreases relative to the mass \(m\) of the spin-2 boson, the soliton's angular frequency approaches \(m\), leading us further into the non-relativistic regime. Consequently, the soliton becomes more diluted, meaning it is less massive and more spatially extended.}

The total soliton number of particles is just given by the expression $M_{\text{sol}} = m N_{\text{sol}}$, since the magnitude of the soliton energy is much less than the soliton mass in the $(\textcolor{black}{\bar\mu}/m) \ll 1$ regime, e.g. spin-2 particles within solitons are cold. From now on, unless stated otherwise, when we refer to solitons, we mean extremally polarized solitons.

\textcolor{black}{As mentioned in the introduction, if dark matter consists of light bosonic particles with spin \(s\) (where \(m \ll\,\text{eV}\)), solitons can form through gravitational condensation within dark matter halos \cite{Jain:2023ojg}. This phenomenon exhibits similar characteristics for scalar, vector, and tensor solitons, as they all share the same radial density profile (they represent ground state solutions of the (2$s$+1) component Schrödinger-Poisson system \cite{Jain:2021pnk, Chen:2023bqy}). Solitons composed of higher-spin fields maintain the same mean number density as scalar solitons under similar initial conditions\,\cite{Jain:2023ojg, Chen:2023bqy}.}

\textcolor{black}{Solitons would nucleate in the first dark matter minihalos at \( z \gtrsim 10 \), with a heavier mass for heavier minihalo masses, $M_{\text{Mh}}$. For the case of scalar solitons, cosmological numerical simulations have found the following core-halo relation: 
$M_{\text{sol}} \propto M_{\text{Mh}}^{\alpha}m^{-3/2(1-\alpha)}(1+z)^{3/4(1-\alpha)},$
where \( \frac{1}{3} \leq \alpha \leq \frac{3}{5} \) \cite{Escudero:2023vgv, Schive:2014hza, Mocz:2017wlg, Nori:2020jzx, Mina:2020eik, Schwabe:2016rze, Chan:2021bja, Zagorac:2022xic}.
 We use this formula for tensor solitons to have a rough idea of the possible range of soliton masses formed in the first dark matter minihalos. We emphasize the need for additional cosmological numerical simulations to further investigate this case. For instance, examining the production of spin-2 dark matter in a specific model, such as that outlined in Refs.\,\cite{Gorji:2023cmz, Gorji:2025tos}.}

\textcolor{black}{For light spin-2 particles, where \( m \ll\,\text{eV}\), we will demonstrate in the following sections that a resonance phenomenon in solitons leads to an exponential production of photons. The signal is highly monochromatic, with a central frequency peak around \( m \). For detection using radio telescopes on Earth, the mass of the spin-2 particle should range as
$10^{-7}\,\text{eV} \lesssim m \lesssim 10^{-3}\,\text{eV}$.
The lower limit corresponds to a signal frequency of about \( 30 \text{ MHz} \). The absorption and scattering of low-frequency photons by the ionosphere make it highly challenging to detect frequencies below this threshold\,\footnote{\textcolor{black}{If we consider projected space-based facilities, the lower mass limit can extend to even smaller values. For example, the Orbiting Low Frequency Antennas for Radio Astronomy Mission (OLFAR) \cite{2020AdSpR..65..856B, 2016ExA....41..271R}, which plans to deploy thousands of nanosatellites on the far side of the Moon, would enable the detection of signals with frequencies down to about \( 0.30 \text{ MHz} \). OLFAR would broaden the parameter space of interest for the spin-2 particle mass to 
$10^{-9}\,\text{eV} \lesssim m \lesssim 10^{-3}\,\text{eV}$. Space-based detection has also been mentioned in other scenarios in the context of dark matter indirect searches. For instance, see Ref.\,\cite{Choi:2022btl} regarding ultra-long radio wave detection for axion-photon conversion during encounters between axion self-similar minihalos and neutron stars.}}.}

\textcolor{black}{For $100 \lesssim z \lesssim 10$, $\alpha=2/5$, and dark matter minihalos masses $ 10^{-4}\,M_{\odot} \lesssim M_{\text{Mh}} \lesssim 10^{8}\,M_{\odot} $\,\cite{Escudero:2023vgv}, we have $ 10^{-14}\,M_{\odot} \lesssim M_{\text{sol}} \lesssim 10^{-5}\,M_{\odot}$, when the light spin-2 boson mass ranges as $10^{-7}\,\text{eV} \lesssim m \lesssim 10^{-3}\,\text{eV}$. Equivalently, through Eq.\,(\ref{eq:Msol}), we have  $ 10^{-16} \lesssim (\bar\mu/m) \lesssim 10^{-4}$. We will later see in Sec.\,\ref{Sec.RC} that this ratio determines the coupling strength necessary for enabling parametric resonance of photons in tensor solitons.} 

\section{Parametric resonance of photons for a tensor field homogeneous condensate}
\label{Sec.PR}

Interactions between the dark tensor and electromagnetic fields can lead to the generation of electromagnetic radiation through parametric resonance. This phenomenon has been explored in scalar \cite{Hertzberg:2018zte, Hertzberg:2020dbk} and vector solitons \cite{Amin:2023imi}. We extend that analysis to the case of our interest, explicitly focusing on the dimension-6 operators listed in Eq.\,(\ref{Eq:Ops}). 

The main idea is that these operators will result in a system of equations of motion for the electromagnetic field in $k$-space, where the dark tensor field can induce a periodically oscillating pump. In the small coupling or amplitude regime—satisfied for our EFT condition \(g^2 \bar{H}^2 \ll 1\)—the system exhibits a spectrum of narrow bands of instability or resonant bands, which are spaced at \((k/m)^2 \approx n^2\), with \(n=1,2,3,\ldots\). Within these resonance bands, the Fourier modes of the electromagnetic field experience exponential growth, represented as \(\Avec_\kvec(t) \propto e^{\mu_\kvec^{(n)} t}\), where \(\mu_\kvec^{(n)}\) is the Floquet exponent for the \(n\)-th resonant band. Consequently, the photon occupancy number increases dramatically as \(n_\kvec(t) \propto e^{2 \mu_\kvec^{(n)} t}\). Since the width of the resonant bands diminishes as the band number increases, the resonance phenomenon is predominantly driven by the first instability band. This characteristic allows us to perform a manageable analytical analysis with results that agree remarkably well with full numerical calculations. 
The radiation produced extracts energy from the \textcolor{black}{dark tensor field}, creating a promising avenue for dark matter detection from Earth.

\subsection{Electromagnetic equation of motion}
\label{sub:EOM}

The electromagnetic field's equation of motion for operators under study is linear in the vector potential $\Avec(t,\xvec)$. We work in the Coulomb gauge so that $\nabla \cdot \Avec = 0$. For a homogeneous configuration, the $k-$space is particularly useful. We define  $k_i = (\kvec)_i$, $H_{ij} = [\Hvec(t)]_{ij}$ and $A_i = [\Avec_\kvec(t)]_i$, where  
\ba{\label{eq:AFT_equation}
\Avec(t,\xvec) = \int \! \frac{\mathrm{d}^3 \kvec}{(2\pi)^3} \, \Avec_\kvec(t) \, e^{i \kvec \cdot \xvec}\,.
}
The Fourier representation of the electromagnetic field's equation of motion takes the closed form
\ba{\label{eq:OPQ_equation}
    \mathbb{O}_{ij} \ddot{A}_j + \mathbb{P}_{ij} \dot{A}_j + \mathbb{Q}_{ij} A_j = 0 \;,
}
where the matrix coefficients are 
\bsa{eq:OPQ_list}{
    \mathbb{O}_{ij} & = \begin{cases}
    \delta_{ij} & \hspace{0.6 cm}\text{, \ $\Lscr_\mathrm{int} = \g \Ocal_1$} \\ 
    \delta_{ij} + \bigl( 2 \g |\Hvec|^2 \bigr) \, \delta_{ij} & \hspace{0.6 cm}\text{, \ $\Lscr_\mathrm{int} = \g \Ocal_2$} \\ 
     \delta_{ij}+g^2\frac{k_i k_m}{k^2}H_{ml}H_{lj} + g^2 \frac{k_ik_l}{k^2}H_{ml}H_{jm} - g^2 H_{ki} H_{jk} - g^2 H_{ik} H_{kj} 
     & \hspace{0.6 cm}\text{, \ $\Lscr_\mathrm{int} = \g \Ocal_3$} \\
    \end{cases} \\ 
    \mathbb{P}_{ij} & = \begin{cases}
    0 & \text{, \ $\Lscr_\mathrm{int} = \g \Ocal_1$} \\ 
    \bigl( 4 \g \Hvec \cdot \dot{\Hvec} \bigr) \, \delta_{ij} & \text{, \ $\Lscr_\mathrm{int} = \g \Ocal_2$} \\ 
    g^2 \frac{k_ik_m}{k^2}\dot{H}_{ml}H_{lj} + g^2\frac{k_ik_m}{k^2}H_{ml}\dot{H}_{lj} + g^2 \frac{k_ik_l}{k^2}\dot{H}_{jm}H_{ml} & \text{, \ $\Lscr_\mathrm{int} = \g \Ocal_3$}\\
    + g^2 \frac{k_lk_i}{k^2}\dot{H}_{ml}H_{jm}-g^2\dot{H}_{ik}H_{kj} - g^2 H_{ik}\dot{H}_{kj}-g^2 \dot{H}_{ki}H_{jk}-g^2 H_{ki}\dot{H}_{jk}\\ 
    \end{cases} \\ 
    \mathbb{Q}_{ij} & = \begin{cases}
    |\kvec|^2 \, \delta_{ij} + \bigl( 4 i \g \Hvec \cdot \dot{\Hvec} \bigr) \, \epsilon_{ijk} k_k &\hspace{1.1 cm} \text{, \ $\Lscr_\mathrm{int} = \g \Ocal_1$} \\ 
    |\kvec|^2 \, \delta_{ij} + \bigl( 2 \g |\kvec|^2 \, |\Hvec|^2 \bigr) \, \delta_{ij} & \hspace{1.1 cm}\text{, \ $\Lscr_\mathrm{int} = \g \Ocal_2$} \\ 
    \bigl(|\kvec|^2 - g^2 k_m k_l H_{mk} H_{kl} - g^2 k_l k_m H_{mk} H_{kl}  \bigr)\delta_{ij}  
    + g^2 k_l k_i H_{lk} H_{kj}&\hspace{1.1 cm} \text{, \ $\Lscr_\mathrm{int} = \g \Ocal_3$} \\
    - g^2 |\kvec|^2 H_{ki}H_{jk} - g^2 |\kvec|^2 H_{kj}H_{ik} + g^2 k_l k_i H_{jk}H_{kl} \\
    \end{cases} 
    \;.
}
We have explicitly written down all different indices arrangements for a general spatial tensor field with components $H_{ij}$, where $|\Hvec|^2 = H_{ij}H^{ij}$ and $\Hvec \cdot \dot\Hvec = H_{ij}\dot H^{ij}$ are understood. However, the above equations can be further simplified for the particular case of our interest, where the dark tensor field soliton is symmetric under index exchange and traceless. We have assumed $g^2 |\Hvec|^2 \ll 1$ to be within the regime in which EFT is valid. This assumption has allowed us to work to leading order in $g^2$. In addition, spatial gradients of the field has been discarded following the resonance analysis performed in scalar~\cite{Hertzberg:2018zte} and vector solitons\,~\cite{Amin:2023imi}, where the resonant modes hold shorter length scales than the typical soliton length scale, i.e. $k \equiv |\kvec| \gg 1/R$.    

\subsection{Homogeneous dark tensor field} 

Before examining the specific radial profile of non-relativistic dark tensor field solitons, we conduct a Floquet analysis for the case of a homogeneous dark tensor field. This approach has proven helpful for scalar solitons \cite{Hertzberg:2018zte} and vector solitons \cite{Amin:2023imi}. The reason is that the photon growth rate associated with the resonance of solitons in these cases is related to that of the homogeneous case. We carry out numerical and analytical Floquet analysis for the five different polarization states of \(\Hvec\) for the operators \(\Ocal_1\), \(\Ocal_2\), and \(\Ocal_3\), to detect the presence of parametric resonance and its properties. The most interesting quantities to be calculated are the maximum Floquet exponent
(real part), which gives the resonance growth rate, and the associated wavenumber and resonance bandwidth. Results are summarized in Table\,\ref{tab:1}. The Floquet analysis is conducted in detail in Sec. 3.2 of Ref.\,\cite{Amin:2023imi} and in Sec.\,3.2.1-3.2.3 of Ref.\,\cite{Amin:2014eta}. The following provides a brief overview of the general procedure for better clarity. 

The equation of motions for the electromagnetic field in $k$-space, as given in Eq.\,(\ref{eq:OPQ_equation}), can be simplified by applying the Coulomb condition, $\kvec \cdot \Avec = 0$. This constraint allows us to reduce the system of three second-order differential equations to a system of two second-order differential equations, or even to a system of four first-order differential equations. Suppose we are interested in solving the Coulomb gauge condition for $A_3$, then
\begin{equation}
\begin{pmatrix}
    \dot{\Avec}(t) \\
    \ddot{\Avec}(t)
\end{pmatrix}
    = 
    \begin{pmatrix}
        \textcolor{white}{xx} {\bf{0}}\,  & \,\,{\bf{1}}\\
         -\tilde{\mathbb{O}}^{-1}\tilde{\mathbb{Q}}\,  &\,\,\tilde{\mathbb{Q}}
    \end{pmatrix} 
    \begin{pmatrix}
        \Avec(t) \\
    \dot\Avec(t)
    \end{pmatrix}
    \;
    \longrightarrow \, \,
   \dot{\bm s}(t) = \tilde{\mathbb{U}}(t)\bm s(t)\,, 
\end{equation}
where $\bm s(t) = (\Avec(t)\,\, \dot\Avec(t))^{t}$,  $\tilde{\mathbb{O}}_{ij}\equiv \mathbb{O}_{ij}-\mathbb{O}_{i3}k_j/k_3$, and similarly for $\tilde{\mathbb{P}}_{ij}$ and $\tilde{\mathbb{Q}}_{ij}$. Due to the coherent oscillating feature of the dark tensor field condensate, we may have a periodic oscillating $\tilde{\mathbb{U}}(t)$ such that $\tilde{\mathbb{U}}(t) = \tilde{\mathbb{U}}(t+T)$, where $T \approx 2\pi/m$ is the condensate period of oscillation. The Floquet Theory ensures
a solution of the form $\bm s(t) = \sum_{s=1}^{4}c_{s,\kvec} {\bf{P}}_{s,\kvec}(t)e^{\mu_{s,\kvec}t}$, with ${\bf{P}}_{s,\kvec}(t) = {\bf{P}}_{s,\kvec}(t+T)$.
For a fixed wavevector $\kvec$, there are four Floquet exponents $\mu_{s,\kvec}$ and four eigenvectors defining a specific polarization of the radiation output. If the real part of any Floquet exponent is finite, then the Floquet Theory predicts an exponential growth for the particular photon occupancy number. For clarity, we define $\mu_{\kvec,\text{max}}$ as the maximum real part of the Floquet exponent for a specific wavevector $\kvec$, and $\mu_{\text{max}}$ as the maximum real part of the Floquet exponent considering all available wavevectors. We numerically calculate Floquet exponents and eigenvectors by evaluating the system solution at the time $T$ starting from suitable initial conditions. Additionally, we verify these results using analytical approximations. In App.\,\ref{app:general}, we perform an analytical general analysis for the system given in Eq.\,(\ref{eq:OPQ_equation}), while we conduct a detailed analysis of particular examples in App.\,\ref{app:details}.

\subsection{Homogeneous linearly polarized dark tensor field (polarization state = $0$)}

\begin{figure}[t]
\centering
  \includegraphics[scale=0.25]{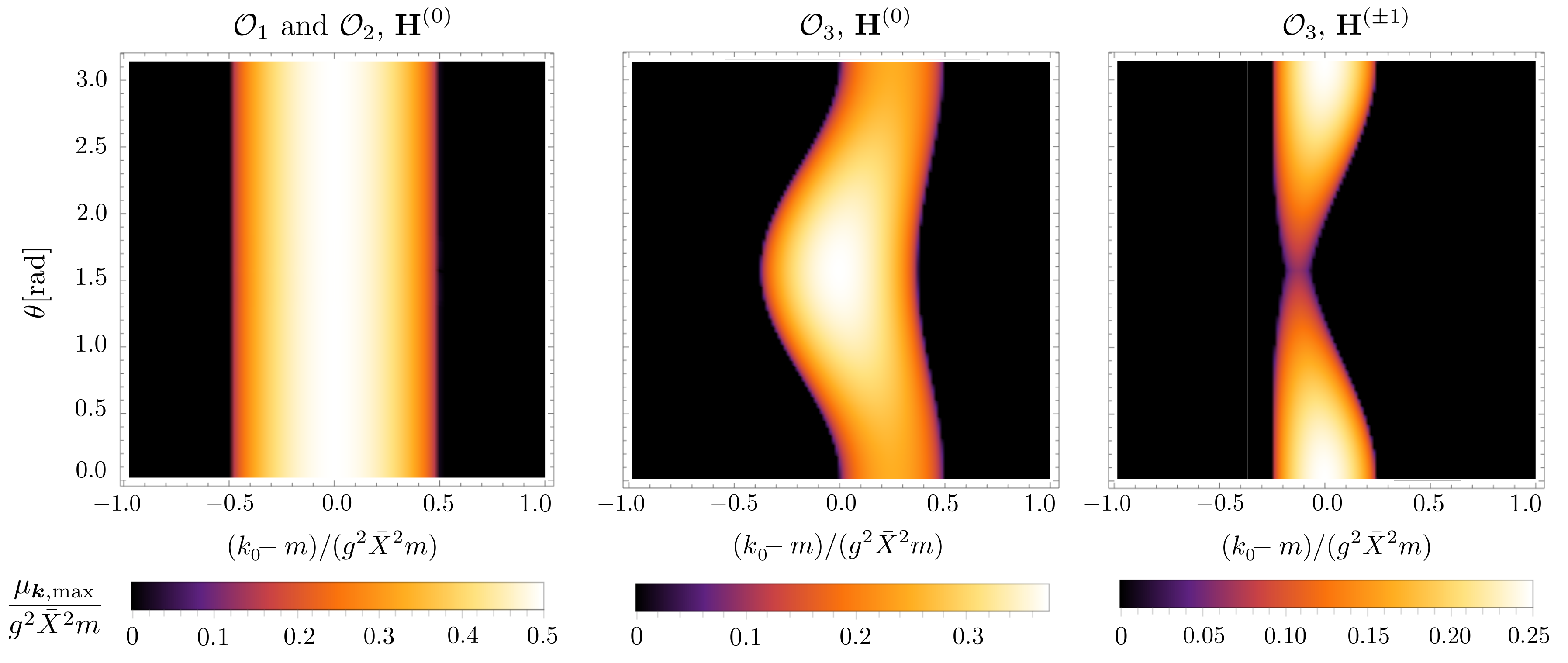}
\caption{\justifying
Parametric resonance phenomenon associated with a homogeneous tensor field
which couples to electromagnetism via several dimension-6 operators. Maximal Floquet exponent $\mu_{\kvec,\text{max}}$ in terms of the shifted and normalized wavenumber $(k-k_0)/(g^2\bar{H}^2m)$  and the angle $\hat{\theta}$ between the direction of propagation of the outgoing radiation and the $\hat{\zvec}$ direction of the tensor soliton. \textit{Left:} Operators $\Ocal_1$ and $\Ocal_2$ for the case of a tensor homogeneous condensate with polarization state equal to $0$. \textit{Middle:} Operator $\Ocal_3$ for the case of a tensor homogeneous condensate with polarization state equal to $0$. \textit{Right:} Operator $\Ocal_3$ for the case of a tensor homogeneous condensate with polarization state equal to $1$.} 
\label{fig:panel}
\end{figure}

We consider a homogeneous dark tensor field with zero polarization state written as
\begin{equation}
\Hvec^{(0)} (t) = \frac{\bar{H}}{\sqrt{6}} 
\begin{bmatrix}
      -\text{cos}(m t)&0&0\\
          0&-\text{cos}(m t)&0\\
          0&0&2\text{cos}(m t)
         \end{bmatrix}\,, 
         \label{eq:Ho}
\end{equation}
where $\bar{H}$ is a constant magnitude,  $\Hvec^{(0)}$ is linearly polarized in each rectangular axis $(\hat{\xvec},\hat{\yvec},\hat{\zvec})$, and we have taken $\omega \approx m$. The Lagrangian density for operators $\Ocal_1$ and $\Ocal_2$ involve the 
contraction between two non-relativistic dark tensor fields,  $H_{ij}H^{ij}$ or $H_{ij}\dot H^{ij}$, i.e., their indices do not mix with those from the electromagnetic field strength tensor or its dual. So, both operators 
produces the same isotropic feature for the radiated photons, and the maximum Floquet exponent (real part) does not depend on the wavevector orientation.  We notice that the isotropic emission is a characteristic of the scalar field homogeneous condensate\,\cite{Hertzberg:2018zte}. Indeed, $|\Hvec(t)|^2 = \bar{H}^2\text{cos}^2(mt)$ can be seen as the square of a scalar field homogeneous condensate, $\phi(t) = \bar{\phi}\,\text{cos}(mt)$, where $\bar{\phi}$ is a constant scalar field amplitude. This isotropic characteristic of the electromagnetic output is also applicable to the vector homogeneous condensate when we make the substitution \( \Hvec \rightarrow \Xvec \) in \( \Ocal_1 \) and \( \Ocal_2 \) \cite{Amin:2023imi}. Here, \( \Xvec(t) = \bar{X} \cos(mt) \hat{\zvec} \) denotes a linearly polarized vector field, with \( \bar{X} \) being a constant amplitude. Numerical results associated with operator $\Ocal_1$ \textcolor{black}{are shown in the left} panel of Fig.\,\ref{fig:panel}, which is indistinguishable from those associated with operator $\Ocal_2$. The independence of the growth rate concerning the polar angle between $\kvec$ and $\Hvec$, i.e. the isotropic nature of the emitted radiation,  produces just a vertical band centered at $ k_0 = m + O(g^4\bar{H}^4m)$ with a maxima growth rate $\mu_{\text{max}} \approx (1/2)g^2\bar{H}^2m$. Details regarding the analytical derivation of the maximum Floquet exponent for $\Ocal_2$ are given in App.\,\ref{apphomo:H002}.    

The fact that operator $\Ocal_3$ mixes indices between the tensor and electromagnetic field strength tensor leads to a richer equation of motion for the vector potential in the $k$-space.  The middle panel of Fig.\,\ref{fig:panel} shows the corresponding Floquet chart indicating a dependence of the maximal Floquet exponent with respect the polar angle ${\bf{\hat{\theta}}}$, i.e. the angle between the wavevector $\kvec$ of the outgoing radiation and the $\hat{\zvec}$ direction of the tensor soliton. The maximal (real part) Floquet exponent expression with an arbitrary polar angle $\hat{\theta}$, the associated wavenumber placed at the center of the dominant resonant band, and its bandwidth take the form (see App.\,\ref{apphomo:H003} for the analytical derivation)
\begin{align}
\mu_{\kvec,\text{max}}(\theta) &\approx g^2 \bar{H}^2 m\frac{\sqrt{51/2}}{16} \left(1 - \frac{20\text{cos}(2\theta)}{51}+\frac{\text{cos}(4\theta)}{51}  \right)^{1/2}\,,\\
k_0 &= \frac{(k_{l,{\rm edge}}+k_{r,{\rm edge}})}{2} = m +  g^2 \bar{H}^2 m \left(\frac{5}{24}+\frac{\text{cos}(2\theta)}{8}\right)  
+ O(g^4\bar{H}^4m)\,,\\
\Delta k &= (k_{r,{\rm edge}}-k_{l,{\rm edge}}) = g^2 \bar{H}^2 m\left(\frac{5}{8}-\frac{\text{cos}(2\theta)}{8} \right) + O(g^4\bar{H}^4m)\,.
\end{align}
In complete agreement with numerical results, the maximum and minimum growth rate for the \textcolor{black}{photon occupancy number} are reached at $\theta = \pi/2$ and $\theta = 0$ and the corresponding Floquet exponents $\mu_{\kvec, \text{max}}(\theta)$ are related as $\mu_{\kvec, \text{max}}(0)/\mu_{\kvec, \text{max}}(\pi/2)=3/4$, where the maxima growth rate $\mu_{\text{max}} \approx (3/8)g^2\bar{H}^2m$ is centered at $k_0 = m + O(g^2\bar{H}^2m)$. The bandwidth of the resonant band is maximal at $\theta = \pi/2$ and minimal at $\theta = 0$, while the central wavenumber approaches $m$ from the right as the Floquet exponent increases. The outgoing radiation at the equator plane
is linearly polarized in the $\hat{\zvec}$ direction.

\subsection{Homogeneous circularly polarized dark tensor field (polarization state = $ 1$)}

We consider a homogeneous dark tensor field with polarization state  $ 1$ as
\begin{equation}
\Hvec^{(+1)} (t) = \frac{\bar{H}}{2} 
\begin{bmatrix}
      0&0&\text{cos}(m t)\\
          0&0&\text{sin}(m t)\\
          \text{cos}(m t)&\text{sin}(m t)&0
         \end{bmatrix}\,. 
         \label{Eq:Hpm1}
\end{equation}
As before, we have taken the oscillation frequency $\omega \approx m $. 
For the operators  $\Ocal_1$ and $\Ocal_2$, the time dependence of the tensor field disappears in Eqs.\,(\ref{eq:OPQ_list}), meaning that the resonance phenomenon does not occur. This is easy to see because of   $|\Hvec^{(+ 1)}(t)|^2 = \bar{H}^2/2$.
In contrast, the operator $\Ocal_3$ leads to a complex time-dependent equation of motion for the Fourier modes of the vector potential through the oscillating tensor field. The emitted radiation has a defined polarization. As shown in the right panel of Fig.\,\ref{fig:panel},  the maximum growth rate for resonance, $\mu_{\text{max}} \approx (1/4)g^2\bar{H}^2m$ placed at $k_0 = m + O(g^2\bar{H}^2m)$, occurs at $\theta = 0$ and $\pi$, while the minimum is found at $\theta = \pi/2$. At both poles, the radiation is left-handed circularly polarized and is emitted perpendicular to the $\hat{\xvec}-\hat{\yvec}$ plane.

\subsection{Homogeneous circularly polarized dark tensor field (polarization state = $2$)}

We consider a homogeneous dark tensor field with a polarization state $  2$ as
\begin{equation}
\Hvec^{(+ 2)} (t) = \frac{\bar{H}}{2} 
\begin{bmatrix}
      \text{cos}(m t)&\text{sin}(m t)&0\\
          \text{sin}(m t)&-\text{cos}(m t)&0\\
          0&0&0
         \end{bmatrix}\,. 
\end{equation}
 The time dependence of the field is absent in Eqs.\,(\ref{eq:OPQ_list}) for $\Ocal_1$, $\Ocal_2$, and $\Ocal_3$. For operators $\Ocal_1$ and $\Ocal_2$ this is easy to see because of   $|\Hvec^{(+ 2)}(t)|^2 = \bar{H}^2/2$. Since the parametric resonance requires an oscillating tensor field with a characteristic oscillation period, no parametric resonance is associated with  $\Hvec^{(\pm 2)}$.

\begin{table}[h]
  \centering
  \begin{tblr}{
      colspec={lllll},
      row{1}={font=\bfseries},
      column{1}={font=\itshape},
      row{even}={bg=gray!10},
    }
          \text{Polarization state}    &\textcolor{white}{xxxxxxxx} $0$  & \textcolor{white}{xxxxxxx}$\pm 1$  &\textcolor{white}{xxx} $\pm 2$    \\
          \text{Features}   &  \textcolor{white}{xxxxx}$\mu_{\text{max}}, k_0, \Delta k$  & \textcolor{white}{xxxxx}$\mu_{\text{max}}, k_0, \Delta k$   & \textcolor{white}{x}$\mu_{\text{max}}, k_0, \Delta k$     \\
          
    \toprule
    \hline
   $\Lscr_\mathrm{int} = g^2\Ocal_1$  & $\frac{g^2\bar{H}^2m}{2}\,,  m \,, g^2\bar{H}^2m$ & \textcolor{white}{xx}$  --------$ & $--------$  \\
    $\Lscr_\mathrm{int} = g^2\Ocal_2$  & $\frac{g^2\bar{H}^2m}{2}\,,  m\,, g^2\bar{H}^2m$ & \textcolor{white}{xx}$  --------$ & $--------$   \\
   $\Lscr_\mathrm{int} = g^2\Ocal_3$  &  $\frac{3g^2\bar{H}^2m}{8}\,, m \,, \frac{3g^2\bar{H}^2m}{4} $ &\,\,\,\, $\frac{g^2\bar{H}^2m}{4}\,, m\,,\frac{g^2\bar{H}^2m}{2}$ &  $--------$\\
    \bottomrule
  \end{tblr}
  \caption{\justifying
  Parameters associated with parametric resonance
  phenomenon for a homogeneous dark tensor field with 
   polarization states $0,\pm 1, \pm 2$, e.g. $\Hvec^{(0)},\Hvec^{(\pm 1)},$ and $\Hvec^{(\pm 2)}$, respectively. All written results correspond to the leading order quantities of the mass $m$ times powers of $\mathcal{O}(g\bar{H})$.}
    \label{tab:1}
\end{table}

\section{Parametric resonance of photons of dark tensor field solitons}
\label{Sec.PRS} 

\subsection{Resonance condition}
\label{Sec.RC}

We provide a detailed analysis of the resonance phenomenon for a homogeneous tensor field condensate in Section \ref{Sec.PR}. Our focus, however, is on tensor solitons and their potential as engines for generating electromagnetic radiation through parametric resonance. Previous studies on scalar solitons \cite{Hertzberg:2020dbk, Hertzberg:2018zte} demonstrated that resonance in these configurations operates similarly to that of a homogeneous condensate, provided that the spatial dimensions of the solitons are sufficiently large.  Authors in \,\cite{Amin:2023imi} adhered to this idea for the case of vector solitons. In our work, we adopt a similar approach due to the universality of the spherically symmetric profile. This profile addresses both the one-component and multi-component Schrödinger-Poisson systems for scalar and higher spin bosonic particles, respectively. \textcolor{black}{Consequently, we propose the following approximation for the maximum growth rate in the context of parametric resonance within a tensor soliton, which we denote as $\mu_{\text{max}}^{\text{(sol)}}$:}
\begin{equation}
\mu_{\text{max}}^{\text{(sol)}}=\left\lbrace\begin{array}{c} \mu_{\text{max}}^{\text{(hom)}} - \mu_{\text{esc}}\,,\hspace{0.2cm}\mu_{\text{max}}^{\text{(hom)}} > \mu_{\text{esc}} \\ \textcolor{white}{xxxxx}0\,,\textcolor{white}{xxxxx}\mu_{\text{max}}^{\text{(hom)}} \lesssim \mu_{\text{esc}} \end{array}\right.\, \label{eq:RCon}   
\end{equation}
where $\mu_{\text{esc}}\approx 2/R_{\text{sol}}$ is the typical escape rate for photons in solitons. Additionally, \(\mu_{\text{max}}^{\text{(hom)}}\) represents the maximum growth rate associated with a tensor homogeneous condensate such that ${\bf H}(t,\bm x) = {\bf H}(t)$. The resonance phenomenon occurs only when this growth rate exceeds the escape rate. \textcolor{black}{The approximation given in Eq.\,(\ref{eq:RCon}) is grounded in the following reasoning: 
Consider substituting the tensor soliton condensate, defined by a certain central amplitude, with a corresponding homogeneous field configuration that also upholds the same amplitude. If the width of the soliton is significantly large, the maximum growth rate for the homogeneous case $\mu_{\text{max}}^{(\text{hom})}$ should closely match the maximum growth rate of the soliton, $\mu_{\text{max}}^{(\text{sol})}$. In contrast, if the spatial extension of the tensor soliton is sufficiently small, meaning that the typical escape rate for photons, denoted as \(\mu_{\text{esc}}\), is higher than \(\mu_{\text{max}}^{(\text{hom})}\), the pairs of photons generated will escape the configuration more quickly than the time needed for the next pair to be created. In this case, Bose-Einstein statistics become ineffective, as there will likely be no more than $O(1)$ photons present in the soliton at any point in time to facilitate exponential growth. As a result, the resonance effect will be highly repressed, resulting in $\mu_{\text{max}}^{(\text{sol})} = 0$.}

To support our claim, we perform a detailed analysis in App.\,\ref{app:details2} of the resonance phenomenon for the case of tensor solitons, where spin-2 particles couple to electromagnetism via the operator  $\mathcal{O}_2=-(1/2) F_{\mu\nu}F^{\mu\nu}H_{\alpha\beta}H^{\alpha\beta}$, operator which takes the form $({\bf{E}} \cdot {\bf{E}})({\bf{H}} \cdot {\bf{H}}) - ({\bf{B}} \cdot {\bf{B}})({\bf{H}} \cdot {\bf{H}})$ in the non-relativistic regime.
Closely following the methodology developed in Ref.\,\cite{Hertzberg:2018zte} for scalar solitons, we 
start from the equation of motion of the electromagnetic vector potential $\Avec(t,{\bf{x}})$  as
\begin{equation}
\ddot{\Avec} - \nabla^2{\Avec} - 2\omega g^2H^2(r)\text{sin}(2\omega t)\dot{\Avec}=0\,,
\end{equation}
where $|{\bf{H}}^{(0)}(t,\xvec)|^2=H^2(r)\text{cos}^2(\omega t)$ from Eq.\,(\ref{Eq:H0}) and we have kept terms until $O(g^2H^2(r))$. We express $\Avec(t,\xvec)$ using a vector spherical harmonic decomposition according to
\begin{equation}
    \Avec(t,{\bf{x}}) = \int_0^{\infty} \frac{dk}{2\pi}\sum_{l=1}^{\infty}\sum_{m=-l}^{l} \left[ v_{klm}(t){\bf{M}}_{lm}(k,{\bm{x}}) - w_{klm}(t){\bf{N}}_{lm}(k,{\bm{x}}) \right]\,,
\end{equation}
where  ${\bf{M}}_{lm}(k,{\bm{x}})$ and ${\bf{N}}_{lm}(k,{\bm{x}})$
are the vector spherical wavefunctions, which depends on the vector spherical harmonics ${\bf{\Phi}}_{lm}({\bm{\hat{x}}}), {\bf{Y}}_{lm}({\bm{\hat{x}}}), {\bf{\Psi}}_{lm}({\bm{\hat{x}}})$, and $v_{klm}(t)$ and $w_{klm}(t)$ are the electromagnetic modes. This decomposition enables us to address the tensor soliton's spatial dependence better. By maintaining the Coulomb gauge, we derive integro-differential equations for the electromagnetic modes in $k$-space as follows:
\begin{align}
\ddot{v}_{klm}(t) + k^2v_{klm}(t) - \omega g^2\text{sin}(2 \omega t)
\int_0^{\infty} \frac{dk'}{\pi} \dot{v}_{k'lm}(t) \widetilde{H^2}( k-k') 
 &=0\,,
 \label{eq:modesvklm}\\
\ddot{w}_{klm}(t) + k^2w_{klm}(t) - \omega g^2\text{sin}(2 \omega t)
\int_0^{\infty} \frac{dk'}{\pi}  \dot{w}_{k'lm}(t)  \widetilde{H^2}( k-k') &=0\,,
\end{align}
where $\widetilde{H^2}( k-k')$ is the Fourier transform of the square of the soliton radial profile. Interestingly enough, electromagnetic modes $v_{klm}(t)$ and $w_{klm}(t)$  do not couple with each other and satisfy each of them the same equation of motion. This is because the original equation in the coordinate space does not mix them, as a rotor would do, which is the case for the operator $\mathcal{O}_1$.

\begin{figure}[!]
\centering
  \includegraphics[scale=0.3]{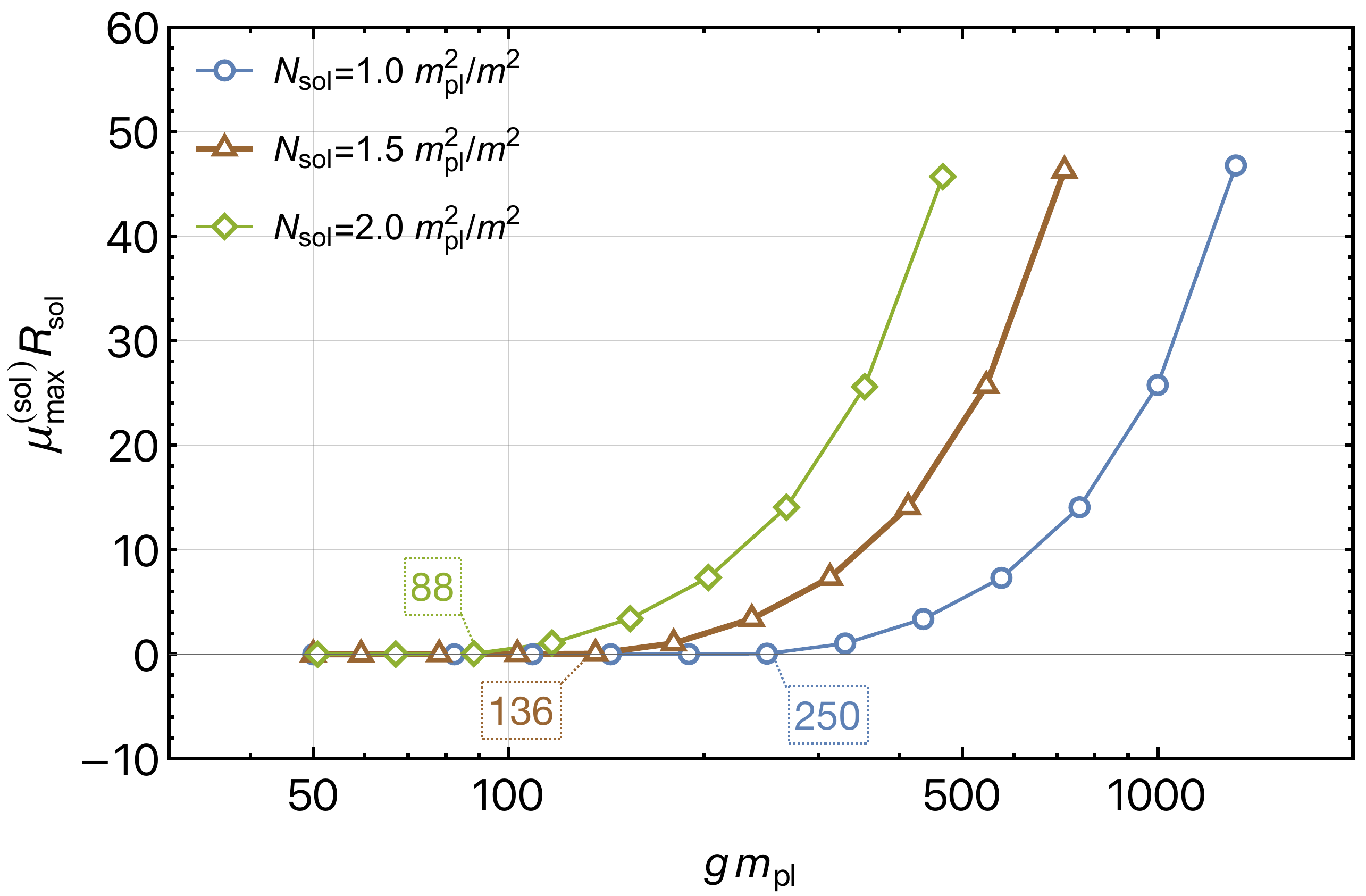}
\caption{\justifying Maximum real part of the soliton Floquet exponent times the soliton radius $\mu^{\text{(sol)}}_{\text{max}}R_{\text{(sol)}}$ in terms of the tensor boson-photon coupling constant $g$ for spherically symmetric tensor solitons with zero polarization state and $N_{\text{sol}}=1.0\,m_{\text{pl}}^2/m^2$, $N_{\text{sol}}=1.5\,m_{\text{pl}}^2/m^2$, and $N_{\text{sol}}=2.0\,m_{\text{pl}}^2/m^2$ (blue, brown and green points, respectively). The plot is obtained by numerically solving Eq.~(\ref{eq:modesvklm}). The \textcolor{black}{approximate values for the coupling constant} (in units of the reduced Planck mass) to trigger the resonance phenomenon are indicated in little colored boxes.}
 \label{SProf}
\end{figure}

\textcolor{black}{Figure\,\ref{SProf} displays the numerical results obtained for the maximum growth rate of electromagnetic radiation in tensor solitons with a zero polarization state. The data is presented for different total particle numbers \( N_{\text{sol}} \): \( 1.0 \, m_{\text{pl}}^2/m^2 \) (blue points), \( 1.5 \, m_{\text{pl}}^2/m^2 \) (brown points), and \( 2.0 \, m_{\text{pl}}^2/m^2 \) (green points). These results are in strong agreement with the proposed resonance condition outlined in Eq. (\ref{eq:RCon}),} resonance occurs when the maximum growth rate of the corresponding homogeneous condensate is comparable to the soliton escape rate. We observe that as the number of particles in the tensor soliton increases (or equivalently, as the soliton mass increases), the coupling strength required to trigger the resonance decreases. This is because a larger soliton mass is associated with a higher central amplitude. 

Using the ``universal'' non-relativistic soliton radial profile which gives expressions for the soliton radius $R_{\text{sol}}$, Eq.\,(\ref{eq:Rsol}), and the soliton central amplitude  $\bar{H}$, Eq.\,(\ref{eq:Hprofile}), together to the homogeneous maximum growth rate  $\mu^{\text{(hom)}}_{\text{max}} \approx (1/2)g^2\bar{H}^2m$ and the escape rate $\mu_{\text{esc}} \approx 2/R_{\text{sol}}$, we may determine the critical value for the coupling constant above which there is resonance, $g_{\text{crit}}$. \textcolor{black}{Since both the soliton central amplitude and the soliton radius depend on the ratio $(\bar\mu/m)$, we anticipate $g_{\text{crit}} = g_{\text{crit}}(\bar\mu/m)$.} However, we must also ensure that the condition $g^2\bar{H}^2\ll 1$, as stated in Eq.\,(\ref{eq:EFTcond}), is satisfied for the validity of our effective field theory framework. Putting it all together, we obtain a range for the coupling constant  
\begin{equation}
 \left( \frac{\textcolor{black}{\bar\mu}}{m} \right)^{-3/4} < gm_{\text{pl}} \ll \left( \frac{\textcolor{black}{\bar\mu}}{m} \right)^{-1}\,,
\label{Rcond}
\end{equation}
where $g_{\text{crit}} \equiv \left( \textcolor{black}{\bar\mu}/m \right)^{-3/4} m_{\text{pl}}^{-1}$ is understood and $\textcolor{black}{\bar\mu}/m \ll 1$ is the typical value for tensor solitons within the non-relativistic regime. \textcolor{black}{As we mentioned before, the parameter space for spin-2 particle masses from the perspective of radio detection on Earth is $10^{-7}\,\text{eV}\lesssim m \lesssim\,10^{-3}\,\text{eV}$. Using the core-halo relation mentioned in Sec.\,\ref{Darktensorpolarizedsolitons} and considering solitons nucleation at $100 \lesssim z \lesssim 10$ in first dark matter minihalos with masses $10^{-4}\,M_{\odot} \lesssim M_{\text{Mh}} \lesssim 10^8\,M_{\odot}$, we have $10^{-16}\,\lesssim \bar\mu/m \lesssim\,10^{-5}$ for  $m \sim\, 10^{-7}\,\text{eV}$.
From the lower bound of Eq.\,(\ref{Rcond}), we obtain $10^{-15}\,\text{GeV}^{-1} \lesssim  g_{\text{crit}} \lesssim  10^{-7}\,\text{GeV}^{-1}$. Similarly, for heavier spin-2 particles with $m \sim\, 10^{-3}\,\text{eV}$, we have 
$10^{-15}\,\lesssim \bar\mu/m \lesssim\,10^{-4}$  leading to $10^{-16}\,\text{GeV}^{-1} \lesssim   g_{\text{crit}} \lesssim  10^{-8}\,\text{GeV}^{-1}$. From now on, we use  $g_{\text{crit}} =  10^{-10}\,\text{GeV}^{-1}$ as a benchmark value for being placed within the ranges above\,\footnote{\textcolor{black}{The same benchmark value was used in Ref.\,\cite{Amin:2023imi} for the case of vector solitons, where dark photons couple to electromagnetism via dimension-6 operators.}}.}

Figure\,\ref{SProf} shows that there is a critical tensor soliton mass corresponding to a given coupling strength. By utilizing the left side of the condition above, along with the ``universal'' formulas for the soliton mass and energy as provided in Eqs.\,(\ref{eq:Msol}) and \,(\ref{eq:Esol}), we can estimate the critical mass and energy. A tensor soliton may undergo resonance if its mass exceeds the critical mass or, equivalently, its energy is below the critical energy. We have
\begin{align}
M_{\text{sol,crit}} &\approx 9 \times 10^{-10} M_{\odot} \left(
\frac{10^{-6}\text{eV}}{m}
\right) \left(
\frac{10^{-10}\text{GeV}^{-1}}{g_{\text{crit}}}
\right)^{2/3}\,,\label{Mcrit}\\
E_{\text{sol,crit}} &\approx -2 \times 10^{-21} M_{\odot} \left(
\frac{10^{-6}\text{eV}}{m}
\right) \left(
\frac{10^{-10}\text{GeV}^{-1}}{g_{\text{crit}}}
\right)^{2}\,.\label{Ecrit}
\end{align}

\subsection{Effective photon mass}

After the formation of tensor solitons, the resonance condition outlined in Eq.\,(\ref{Rcond}) determines whether these objects will experience parametric resonance of photons. It is important to note that our analysis thus far has been conducted assuming massless photons. However, photons can acquire mass due to available free charge in the environment. The square of the photon plasma frequency is proportional to the number of free electrons, \( n_e \), according to   
\begin{equation}
\omega_p^2(z) = \frac{4\pi \alpha_{\text{EM}}n_e(z)}{m_e}\,,
\label{eq:wpz}
\end{equation}
where $m_e$ is the electron mass, $\alpha_{\text{EM}}$ is the fine-structure constant, and $z$ is the redshift. \textcolor{black}{In the very early Universe, the plasma frequency is significantly high due to the large number density of free electrons, being, for instance, $\omega_p \sim 0.1T$ for a temperature $T > m_e$.  As the Universe evolves into the late stages, the number density of free electrons decreases with the redshift. Consider the nucleation of tensor solitons via gravitational condensation\,\cite{Jain:2023ojg} in first DM minihalos at around
 $100 \lesssim z \lesssim 10$. The plasma frequency from the average electron density in the Universe at those redshifts is the order of $O(10^{-14})\,\text{eV}$\,\cite{Escudero:2023vgv}. In the parameter space of our interest for light spin-2 boson masses, $10^{-7}\,\text{eV}\lesssim m  \lesssim 10^{-3}\,\text{eV}$, we see that \( m > \omega_p(z) \), meaning the resonance phenomenon is kinematically feasible\,\footnote{After forming in dark matter halos, tensor solitons may evaporate, potentially reionizing the intergalactic medium, which could impose constraints on the coupling constant and tensor mass
based on current data regarding the cosmic microwave background optical depth. This 
scenario was explored for axion solitons in Ref.\,\cite{Escudero:2023vgv}. While this analysis goes beyond the scope of the present work, we mention it here to highlight the possibility for future discussion.}.} 

Tensor solitons are moving in galactic halos, which, in principle, could jeopardize the resonance phenomenon for breaking the periodicity of the tensor field oscillation in the Fourier representation of the electromagnetic field equation of motion. We can express the plasma frequency as $\omega_p(t) \approx \omega_p g(t)$, where   $g(t) $ is a non-periodic time-dependent function of $O(1)$ that represents the spatial inhomogeneity of free electron density. For instance, the modified equation of motion for the Fourier electromagnetic modes, in the case of the operator $\mathcal{O}_2$ for a homogenous dark tensor soliton that is linearly polarized, Eq.\,(\ref{eq:Ho}), is expressed as (see Eq.\,\ref{eq:H002})  
\begin{equation}
\ddot{\Avec} - 2g^2\bar{H}^2 m\, \text{sin}(2 m t)\dot{\Avec} + \left(k^2 + \omega_p(t)^2\right) \Avec = 0\,.
\end{equation}
The resonance is dominated by the first instability band, so that $k \approx m$ and
$ |\dot\Avec| \sim m|\Avec|$. Requiring that the ratio of the non-periodic plasma term to the periodic tensor field term needs to be small, we have  
\begin{equation}
\left(\frac{\omega_p^2|\Avec|}{2g^2\bar{H}^2m|\dot\Avec|}\right) \sim 2 \times 10^{-6} \left(\frac{\omega_p}{6.4 \times 10^{-12}\,\text{eV}}\right)^2 \left(\frac{10^{-6}\,\text{eV}}{m}\right)^2 \left(\frac{6.6 \times 10^{-12}}{\textcolor{black}{\bar\mu}/m}\right)^{1/2} \,,
\end{equation}
where we have taken $n_e \sim 0.03\, \text{cm}^{-3}$\,\cite{DCARLSON1994431}  for the current average of the electron number density in the interstellar medium and used $g = g_{\text{crit}}$, where $g_{\text{crit}} = 10^{-10}\text{GeV}^{-1}$. The effective photon mass corrections are expected to be negligible for our fiducial parameters. 

\subsection{Tensor soliton mass pile-up}

Equation (\ref{Mcrit}) indicates that there should be an upper limit on the current mass of a subgroup of tensor solitons present today in galactic halos. The solitons within this subgroup are those capable of undergoing parametric resonance after formation, which depends on the coupling between the tensor field and electromagnetism, as well as the polarization states of the solitons. However, depending on the ultraviolet completion theory related to dark tensor fields, multiple dimension-6 operators could simultaneously induce resonance.
Additionally, the energy of the solitons is degenerate with respect to the polarization states in the absence of self-interactions. This means we should not expect to observe a bias towards any particular states due to stability concerns. Furthermore, we have not provided a comprehensive list of all possible dimension-6 operators available for dark tensor-photon coupling, opening up even more resonance possibilities. Consequently, it is reasonable to conclude that Equation (\ref{Mcrit}) sets an upper limit on the current mass of most of the tensor soliton population. 

After formation, tensor solitons having an initial total mass $M_{\text{sol}}$ such that $M_{\text{sol}} > M_{\text{sol,crit}}$ would radiate the excess of particles via resonance until $M_{\text{sol}} \rightarrow M_{\text{sol,crit}}$. Based on that, we should expect to have a sizeable sub-population of tensor solitons with masses $M_{\text{sol}}\approx M_{\text{sol,crit}}$ in galactic halos today (a similar idea was discussed in Refs.\,\cite{Hertzberg:2018zte} and \cite{Amin:2023imi} for the case of scalar and vector solitons, respectively).  From this point forward, we will consider this sizeable sub-population as the entire soliton population for practical purposes. 

\subsection{Merger condition for tensor soliton collisions}

Although we conclude that today, there should be a significant sub-population of tensor solitons with masses $M_{\text{sol}}\lesssim M_{\text{sol,crit}}$, these compact objects may still have the potential to acquire supercritical masses through collisions and subsequent mergers. In Ref.\,\cite{Hertzberg:2020dbk}, the authors conducted full three-dimensional simulations to analyze the merger conditions of pairs of spherically symmetric axion condensates, including the axion self-interaction. They confirmed the ``golden rule'' observed in simulations of fuzzy dark matter halos\,\cite{Schwabe:2016rze}. Excess particles are radiated away through the emission of scalar waves, resulting in a newly formed spherically symmetric axion condensate with a mass of approximately $70\%$ of the total mass of the progenitor condensates.

This work was expanded in Ref.\,\cite{Amin:2022pzv} to include vector solitons, reinforcing the previously mentioned golden rule. However, the intrinsic spin of the spin-1 dark matter particles in vector solitons leads to a complex evolution of the spin density in the final configuration following a merger. Importantly, while the total spin angular momentum is conserved, the resultant soliton does not necessarily retain all of the initial spin angular momentum. It was observed that the collision and merger of two linearly polarized solitons (with zero initial total spin angular momentum) could result in forming a final core with non-zero spin angular momentum, surrounded by a halo with opposite spin angular momentum. In cases involving multiple mergers of linearly polarized solitons, the central core exhibited a combination of circular and linear polarization, leading to a total non-zero spin angular momentum. Given the similarities between vector and tensor solitons, we expect these observations will also be relevant for tensor soliton mergers, particularly concerning the golden rule and the formation of fractionally polarized solitons after the merger.

It is essential to distinguish between soliton collisions and mergers, as not all collisions result in a final merger. \textcolor{black}{As noted in Refs. \cite{Hertzberg:2020dbk, PhysRevD.74.063504, Gonzalez:2011yg, PhysRevA.54.2185, PhysRevLett.78.4143} regarding scalar solitons,} the system's initial total energy influences the merger process. For a merger to occur, the initial total energy must be negative ($E_{\text{initial}} < 0$), indicating that the system is bound. On the other hand, if the initial total energy is positive, the result will be the passage of the solitons through one another without leading to a merger.

When the initial separation between the solitons is sufficiently large compared to their individual radii, the initial total energy of the system can be estimated by summing their individual total energies, their individual kinetic energies, and the gravitational attraction between the solitons, which can be treated as point masses. Consider a head-on collision between tensor solitons with their individual centers of mass initially separated by a distance $d$ and having masses and energies  $(M_{\text{sol,1}},M_{\text{sol,2}})\approx M_{\text{sol}}\lesssim 
M_{\text{sol,crit}}$ and $(E_{\text{sol,1}},E_{\text{sol,2}})\approx E_{\text{sol}}\lesssim 
E_{\text{sol,crit}}$, respectively, where $M_{\text{sol,crit}}$ and 
$E_{\text{sol,crit}}$ are given by Eqs.\,(\ref{Mcrit}) and  (\ref{Ecrit}). \textcolor{black}{Then, the initial energy of the system is approximated to be
\begin{equation}
E_{\text{initial}} = 2\times E_{\text{sol}} + 2\times \frac{M_{\text{sol}}v^{2}_{\text{cm}}}{2} - \frac{M_{\text{sol}}^{2}}{8\pi m^{2}_{\text{pl}}  d}\,.
\label{Energyinitial1}
\end{equation}
Imposing the condition $E_{\text{initial}} \lesssim 0$ and the ``universal'' expressions for soliton mass and energy, Eqs.\,(\ref{eq:Msol}) and \,(\ref{eq:Esol}), we estimate the condition for soliton merger after collisions as
\begin{equation}
\left(\frac{\textcolor{black}{\bar\mu}}{m}\right) \gtrsim 4 \times 10^{-7}\,\left(\frac{v_{\text{rel}}}{\sqrt{2}\times 220\, \text{km/s} }\right)^2 \,,
\label{Energyinitial}
\end{equation}}
\hspace{-0.32 cm} where $v_{\text{cm}}$ is the velocity of each soliton with respect to the system center of mass, $v_{\text{rel}} = 2 v_{\text{cm}}$ is the relative velocity between solitons, and we have taken $d = 16 R_{\text{sol}}$, which is about 11 times larger than the geometrical mean of the length scales, $\sqrt{2}R_{\text{sol}}$. At such initial distance, solitons are far away enough to justify the expression in Eq.\,(\ref{Energyinitial}). For a comparable treatment regarding scalar solitons, please refer to Ref. \cite{Hertzberg:2020dbk}.   

\subsection{Tensor soliton merger rate in galaxies}
\label{Sec:mergerrate}

We should expect to currently find a fraction of the dark matter, $f_{\text{DM}}$, in the form of tensor solitons within galactic dark matter halos holding a mass just below the critical one, $M_{\text{sol}} \simeq M_{\text{sol,crit}}$. The idea is to estimate the merger rate for solitons, which is necessary to predict subsequent burst emission of photons. The merger rate takes the form
\begin{equation}
\Gamma_{\text{merg}} = \frac{1}{2} \int_{4r_{\text{sch}}}^{r_\text{vir}} 4\pi r^2
\left(  \frac{\rho_{\text{halo}}(r)f_{\text{DM}}}{M_{\text{sol}}} \right)^2
\langle \sigma_{\text{eff}}(v_{\text{rel}})v_{\text{rel}}  \rangle_{\text{merg}} \text{d}r\,,
\label{MR}
\end{equation}
where $r_{\text{sch}}$ is the Schwarzschild radius, $\rho_{\text{halo}}(r)$ is the dark matter density profile within galactic halos, $r$ is the galactic radius,  $v_{\text{rel}}$ is the relative velocity between solitons, $\sigma_{\text{eff}}$ is the effective cross section, $\langle ... \rangle_{\text{merg}}$ means an average over the inside quantities, and $r_\text{vir}$ is the dark matter halo virial radius. The effective cross section, $\sigma_{\text{eff}}$, corresponds to the geometrical cross section enhanced by the gravitational pulling between solitons according to\,(see page 627 in \cite{2008gady.book.....B})
\begin{align}
\sigma_{\text{eff}}(v_{\text{rel}})& = \pi \left( R_{\text{sol}} + R_{\text{sol}}  \right)^2
\left[ 1 + \left(\frac{v^{\text{mutual}}_{\text{esc}} }{v_{\text{rel}} }\right)^2\right] = 4\pi R^2_{\text{sol}} \left( 1 +   \frac{M_{\text{sol}}}{4\pi m^2_{\text{pl}}R_{\text{sol}}v^2_{\text{rel}}} \right)\,, \\
&\approx 125\, m^{-2}(\textcolor{black}{\bar\mu}/m)^{-1}\left[  1 + 10^{6}(\textcolor{black}{\bar\mu}/m)\left(\frac{\sqrt{2}\times 220\,\text{km/s}}{v_{\text{rel}}}\right)^2 \right]\,,
\label{sigmaeff}
\end{align}
where $v^{\text{mutual}}_{\text{esc}} = \sqrt{2(M_{\text{sol}}+M_{\text{sol}})/[8\pi m^2_{\text{pl}}(R_{\text{sol}}+R_{\text{sol}})]}$ is the mutual escape velocity between solitons. We take as a fair approximation that the relative velocity between solitons follows a Maxwell-Boltzmann 3-d isotropic velocity distribution that peaks at  $\sigma_{\text{rel}}$. We estimate the average over the relative velocity of the effective cross section as 
\begin{equation}
\langle \sigma_{\text{eff}}(v_{\text{rel}})v_{\text{rel}}  \rangle_{\text{merg}} = \int_0^{\text{min}(2v_{\text{esc}},\,2 \times 0.85\sqrt{\textcolor{black}{\bar\mu}/m})} 4 \pi \sigma_{\text{eff}}(v_{\text{rel}})v_{\text{rel}} P({v_{\text{rel}}})v^2_{\text{rel}}dv_{\text{rel}}\,,
\label{eq:averagevrel}
\end{equation}
where the spatial dependent escape velocity for solitons in dark halos is calculated by using $v_{\text{esc}} = \sqrt{M_{\text{gal}}(r)/(4\pi m^2_{\text{pl}} r)}$, where $M_{\text{gal}}(r)$ is the galactic mass at the spherical coordinate $r$, and
$P({v_{\text{rel}}})v^2_{\text{rel}}dv_{\text{rel}} \equiv (3/(2\pi\sigma^2_{\text{rel}}))^{3/2}\text{exp}(-3v^2_{\text{rel}}/(2\sigma^2_{\text{rel}}))v^2_{\text{rel}}dv_{\text{rel}}$ measures how much probable is to find two solitons having a relative speed in the interval $(v_{\text{rel}} + dv_{\text{rel}})$.  The upper limit of integration in Eq.\,(\ref{eq:averagevrel}) ensures that collisions lead to mergers by satisfying Eq.\,(\ref{Energyinitial}).

We use the Milky Way Galaxy as a concrete example to estimate the magnitude of the tensor soliton merger rate. In particular, we take the BCKM Galactic model\,\cite{PhysRevD.87.083525}  obtained by means of a
Markov Chain Monte Carlo analysis based on Galactic rotation
curve data. This model considers a dark matter halo holding a Navarro-French-White profile\,\cite{1996ApJ...462..563N}  and a spheroidal bulge overlapped with an axisymmetric disk. We also include Sagittarius A*, the supermassive black hole placed at the Galactic center, whose effects are important within its gravitational influence radius\,\cite{1972ApJ...178..371P}. We compute the relative velocity distributions between tensor solitons following the methodology developed in Ref.\,\cite{Amaral:2023ekd}, where Fast Radio Burst signals\,\cite{Katz:2018xiu} related to primordial black hole-neutron star interactions are studied.  In particular, we use the Eddington inversion formula and a pool of Monte Carlo simulations for the outer Galactic regions, 
fitting our numerical results to the closest Maxwell-Boltzmann 3-d isotropic relative velocity distribution. We apply the Jeans equation and reasonable assumptions for the inner Galactic regions. The BCKM Galactic model and procedure are detailed in Appendix \ref{App:GalacticModel}.

\begin{figure}[!]
\centering
  \includegraphics[scale=0.35]{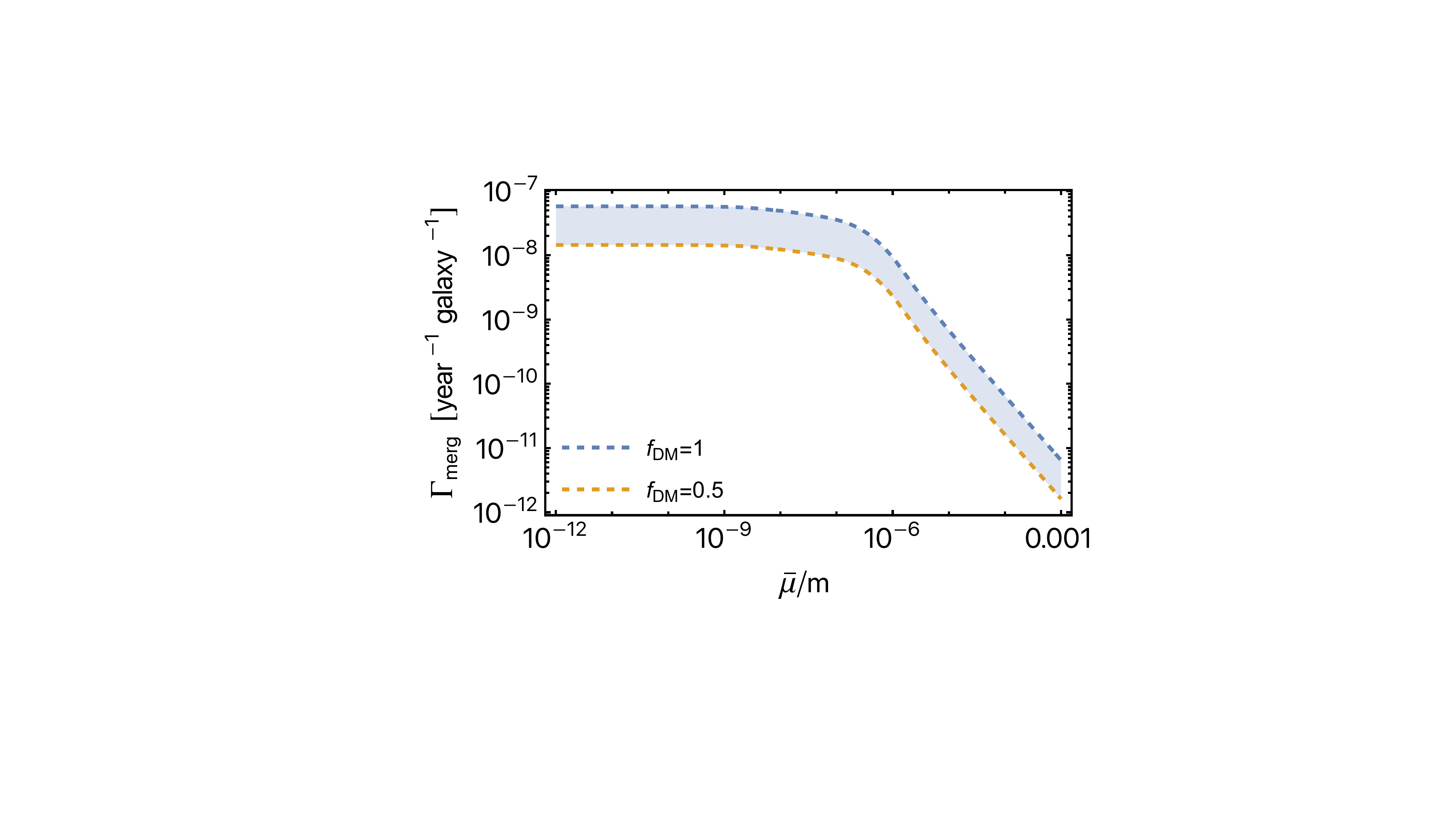}
\caption{\justifying Tensor soliton merger rate in the Galaxy as a function of $\textcolor{black}{\bar\mu}/m$ for a dark tensor mass of $m=10^{-6}\,\text{eV}$ and a dark matter fraction in solitons of $100\%$ and $50\%$.}
 \label{Fig.MRate}
\end{figure}

Figure\,\ref{Fig.MRate} shows an estimate for the tensor soliton merger rate in the Galaxy in terms of the ratio $\textcolor{black}{\bar\mu}/m$ considering a dark matter fraction in solitons of $100\%$ (blue dashed line) and $50\%$ (orange dashed line).  We adopt a phenomenological perspective by treating the ratio $\textcolor{black}{\bar\mu}/m$ and $f_{\text{DM}}$ as free parameters. We notice that there exists a change in the merger rate behavior around $\textcolor{black}{\bar\mu}/m \sim 5 \times 10^{-7}$. At that value, for typical soliton relative velocities, the merger condition in Eq.\,(\ref{Energyinitial}) is barely satisfied and the gravitational focusing closely balances the geometrical cross section in Eq.\,(\ref{sigmaeff}). When $\textcolor{black}{\bar\mu}/m \lesssim 10^{-7}$, the average over the relative velocity of the effective cross section, Eq.\,(\ref{eq:averagevrel}), scales as $v^4_{\text{rel}} \propto (\textcolor{black}{\bar\mu}/m)^2$ and the effective cross section is dominated by the geometrical one so that it scales as  $R^2_{\text{sol}} \propto (\textcolor{black}{\bar\mu}/m)^{-1}$. Considering that the square of the soliton number density scales as $M^{-2}_{\text{sol}} \propto (\textcolor{black}{\bar\mu}/m)^{-1}$,  we have that the merger rate in this regime scales as $(\textcolor{black}{\bar\mu}/m)^{0}$ explaining the flat part of the curve in Fig.\,\ref{Fig.MRate}. When  
$\textcolor{black}{\bar\mu}/m \gtrsim 10^{-6}$, the upper limit integration in 
Eq.\,(\ref{eq:averagevrel}) is dominated by the escape velocity, and the effective cross section in Eq.\,(\ref{sigmaeff}) is dominated by the gravitational focusing. Thus, the merger rate scales as $(\textcolor{black}{\bar\mu}/m)^{-1}$, which explains the decreasing part of the curve in Fig.\,\ref{Fig.MRate}.

\subsection{Electromagnetic signal features}

From now on, suppose after tensor soliton merger, the resultant condensate holds a supercritical mass $M > M_{\text{crit}}=m N_{\text{crit}}$ and undergoes parametric resonance of photons. The resonance leads to an exponential growth of the photon occupancy number, so classical electromagnetic waves constitute the physical output. To characterize the astrophysical signature associated with this phenomenon, we need to estimate the duration of the signal,  the central wavelength, and
the signal bandwidth. 

The time scale linked to the exponential growth of the photon occupancy number should be the order of the inverse of the maximum growth rate. For an estimate, we take  
 $\mu^{(\text{sol})}_{\text{max}} \sim \mu^{(\text{hom})}_{\text{max}}$. Therefore, the ``universal" growth time scale is estimated as
\begin{equation}
\tau_{\text{inst}} = \frac{1}{\mu^{(\text{sol})}_{\text{max}}} \sim \left(\frac{g^2\,m\,\bar{H}^2}{2}\right)^{-1}  \sim 120\,\mu\text{s} \left(\frac{m}{10^{-6}\,\text{eV}}\right)^{-1}
\left(\frac{g_{\text{crit}}}{10^{-10}\,\text{GeV}^{-1}}\right)^{2/3}\,,
\end{equation}
where we have used the resonant condition, Eq.\,(\ref{Rcond}), and the ``universal" soliton central amplitude. This estimate considers an instantaneous burst occurring after the formation of a soliton with supercritical mass, resulting from the merger of the original solitons.

Taking into account that the resonance phenomenon is dominated by the first instability band in the small amplitude or coupling analysis, \textcolor{black}{the signal output should be peaked} at the resonant wavelength
$\lambda_{0} \approx 2\pi/k_0$ with a bandwidth  $\Delta  \lambda_0 = |-2\pi/k_0^2|\Delta k$, where $k_0 \approx m$ and $\Delta k \approx g^2\bar{H}^2m$.
Similarly, we may calculate the peak and the bandwidth in the frequency spectrum, e.g. $\nu_{0}$ and $\Delta \nu_{0}$ respectively, recalling that $\nu_{0}\lambda_{0}=c$. 
Therefore, we have
\begin{align}
\lambda_{0} &\sim 1\,\text{m}\,\left(\frac{10^{-6}\,\text{eV}}{m}\right)\,,\textcolor{white}{xxx}
\Delta \lambda_{0}  \sim 10^{-5}\,\text{m}\left(\frac{10^{-6}\,\text{eV}}{m}\right)\left(\frac{10^{-10}\,\text{GeV}^{-1}}{g_{\text{crit}}}\right)^{2/3}\,, \\
\nu_{0} &\sim 240\,\text{MHz} \left(\frac{10^{-6}\,\text{eV}}{m}\right)^{-1}\,,\textcolor{white}{xxx} \Delta\nu_{0}\sim 3\,\text{kHz}
\left(\frac{10^{-6}\,\text{eV}}{m}\right)^{-1}\left(\frac{10^{-10}\,\text{GeV}^{-1}}{g_{\text{crit}}}\right)^{2/3}\,,
\label{eq:Deltanu0}
\end{align}
where we have used again Eq.\,(\ref{Rcond}), and the ``zero" subscript means that wavelengths and frequencies are measured at the soliton rest frame. On Earth, the signal wavelength undergoes redshifting due to the cosmic expansion as $\lambda = \lambda_0 (1+z)$, where $z$ indicates redshift as usual. Assuming an FRW expanding Universe with a null spatial curvature, if the soliton is located from us at a luminosity distance $d_L = (1+z)f(z)$, where $f(z)=(H_0/c)^{-1}\int_0^{z}(H(z')/H_0)^{-1}dz'$\,\cite{Weinberg:1972kfs}, then the measured spectral flux density on the Earth, i.e. the flux per bandwidth, can be estimated as
\begin{align}
S_{\text{B}} &\sim \frac{(M_{\text{sol}}-M_{\text{sol, crit}})/\tau}{4\pi\Delta\nu (1+z)\,d_L^2}
\sim  \frac{6\, m^{4/3}_{\text{pl}}}{m \,g^{2/3}_{\text{crit}}d_L^2(z)}\,,\label{Eq:UpperSB2}\\
& \sim 4 \times 10^{-13}\, \text{erg cm}^{-2}\text{Hz }^{-1}\text{s}^{-1} \left( \frac{10^{-6}\,\text{eV}}{m} \right)\left( \frac{10^{-10}\,\text{GeV}^{-1}}{g_{\text{crit}}} \right)^{2/3}\left( \frac{1.53}{(H_0/c)d_L(z=1)} \right)^{2}\,,
\label{Eq:UpperSB}
\end{align}
where we have taken $H_0 = 67.66 \times 10^{3}\, \text{m}\, \text{s}^{-1} \text{Mpc}^{-1}$, $\Omega_{\text{m},0} = 0.3111$, $\Omega_{\Lambda,0} = 0.6889$ for a flat $\Lambda$CDM\,\cite{Planck:2018vyg}, an isotropic emission, and $\Delta \nu_0 = \Delta \nu  (1+z)$ given by Eq.\,(\ref{eq:Deltanu0}). 

\textcolor{black}{The resultant soliton mass after a merger would align with the ``golden rule" noted in soliton collision numerical simulations\,\cite{Schwabe:2016rze, Hertzberg:2020dbk}. This rule states that the resultant soliton mass after the merger is approximately $70\%$ of the total mass of the progenitor condensates. If we denote the colliding soliton masses as $(M_{\text{sol,1}}, M_{\text{sol,2}}) \approx M_{\text{sol,crit}}$, the final soliton mass following the merger can be approximated as $ M_{\text{sol}} \approx 0.7(M_{\text{sol,crit}}+M_{\text{sol,crit}})$. Since this resultant mass exceeds the critical soliton mass, as given in Eq.\,(\ref{Mcrit}), resonance is activated, leading to an exponential production of photons. As these photons are emitted, the mass of the soliton diminishes until it reaches $M_{\text{sol,crit}}$, suppressing the resonance phenomenon. Consequently, the liberated energy can be estimated by $(M_{\text{sol}} - M_{\text{sol,crit}}) \approx 0.4 M_{\text{sol,crit}}$, as used in Eq.\,(\ref{Eq:UpperSB2}). }

 The spectral flux density shown in Eq.\,(\ref{Eq:UpperSB}) must be compared with the sensitivities of typical ground-based telescopes in the radio spectrum. For a signal in the radio band of the electromagnetic spectrum, telescopes show sensitivities at around $O(100)\,\mu\text{Jy}$ at $100\,\text{kHz}$ and resolution bandwidth of $O(\text{kHz})$, such as the Square Kilometre Array and the Green Bank telescope\,\cite{SKA1}, being both able to detect polarized light. Here $\text{Jy}$ means Jansky and corresponds to the usual units of the spectral flux density ($\text{Jy} = 10^{-23}\,\text{erg cm}^{-2}\text{Hz }^{-1}\text{s}^{-1}$).
 
 If an extragalactic or Galactic tensor soliton explodes in electromagnetic bursts, the signal strength should not be of concern from the detection point of view. Here we should point out that the spectral flux density shown in Eq.\,(\ref{Eq:UpperSB}) needs to be seen as an upper bound for the signal since we are considering the shortest time scale (instantaneous burst emission) and the maximum converted mass ($\sim M_{\text{sol, crit}}$). Probably, the real situation is slightly different. From the numerical simulations performed in Ref.\,\cite{Hertzberg:2020dbk} for the case of scalar solitons, we know that mergers show three steps: collision, coalescence, and system settling down to the minimum energy configuration (ground state). The releasing particles period is not instantaneous but is on the order of years for the most attractive QCD axion parameters. Since we are looking for configurations where the typical soliton length scale is much larger than the typical resonance time scale, we expect that radiation seeps out from tensor solitons via weaker bursts as the whole configuration settles down to the final ground state condensate. 

\section{Comparison with scalar and vector solitons}

This section discusses the main similarities and differences between tensor solitons and vector and scalar solitons concerning the resonance phenomenon. Knowing similar features for this phenomenon will help us look for a generic signal from these astrophysical objects in galactic halos. Knowing differences will help us look for special features in the signal, aiming to recognize the particle physics nature of the resonant soliton.

\subsection{Similarities among solitons}

\begin{itemize}
    \item No matter the spin multiplicity of the light boson particle constituting the soliton, all solitons share a common radial profile since the non-relativistic behavior of fields is described by a $(2s + 1)$ component Schrödinger-Poisson system\,\cite{Jain:2021pnk}. 

    \item Since all solitons correspond to an oscillating coherent field,
    they may undergo parametric resonance of photons, corresponding to an exponential growth of the photon occupancy number and producing a classical electromagnetic wave output. 

    \item When the resonance is possible for homogeneous condensates of spin-$s$ boson particles, all solitons undergo resonance when the maximum growth rate for the corresponding homogeneous condensate, having an amplitude equal to the soliton central amplitude, is greater than the typical photon light-crossing time, e.g. $\mu^{(\text{hom})}_{\text{max}} \gtrsim \mu_{\text{esc}}$.

    \item When the resonance is possible for homogeneous condensates of spin-$s$ boson particles, a critical soliton mass, or minimum coupling constant, exists above which the resonance phenomenon is activated.

    \item  \textcolor{black}{The spin-2 particle mass range of interest for radio detection on Earth reads as
    $10^{-7}\,\text{eV}\lesssim  m \lesssim10^{-3}\,\text{eV}$. Considering gravitational condensation of tensor solitons\,\cite{Jain:2023ojg} in the earliest dark matter minihalos, particularly around the redshift range of $100\lesssim z \lesssim 10$ and employing the core-halo relation outlined in Sec.\,\ref{Darktensorpolarizedsolitons}, we have estimated the typical values for $g_{\text{crit}}$ needed to enable resonance associated with the typical nucleated soliton masses. When all of this is taken into account, Eq.\,(\ref{Eq:UpperSB}) indicates that the upper bound for the predicted signal strength far exceeds what is necessary for detection using ground-based facilities.}

    \item \textcolor{black}{Since scalar, vector, and tensor solitons share the same radial profile \cite{Jain:2021pnk}, the condition for merger stated in Eq.\,(\ref{Energyinitial})  holds true for all these configurations. Consequently, the merger rate in the Galaxy can be assumed to be applicable to scalar, vector, and tensor solitons. Generally speaking, the total merger rate for solitons in the Galaxy remains low mainly because colliding solitons typically have too much energy to enable a merger. However, we have estimated a maximum rate in the Galaxy of about \(O(10^{-8})\), which is approximately an order of magnitude greater than the rate estimated for scalar solitons in Ref.\,\cite{Hertzberg:2020dbk}. This discrepancy arises from significant improvements in numerical calculations compared to those in Ref.\,\cite{Hertzberg:2020dbk}. The previous analysis assumed that collisions between solitons followed a Maxwell-Boltzmann relative velocity distribution peaked at a constant circular velocity at the solar position. In this work, we have employed the Eddington formalism, along with a series of Monte Carlo simulations, in the outer regions of the Galaxy, while utilizing the Jeans equation in the inner regions, where Sagittarius A* exerts a considerable gravitational influence. These enhancements enable a more precise estimation of the relative velocities between solitons in the Galaxy, which directly affect the merger rate implied by the condition in Eq.\,(\ref{Energyinitial}). In contrast to the approach in Ref.\,\cite{Hertzberg:2020dbk}, where Maxwell-Boltzmann relative velocity distributions are fixed at a constant circular velocity of around \(10^{-3}c\) throughout the Galaxy, our results indicate a radial dependence of relative dispersion velocities as shown in Fig.\,\ref{Numericalfvrelsigmarel} (bottom panel). In the inner Galactic regions, these dispersion velocities can reach values near \(10^{-4}c\). Such smaller relative velocities significantly influence Eq.\,(\ref{eq:averagevrel}), particularly through the upper limit of the integral about the average over the relative velocity of the effective cross section.}
    
\end{itemize}

\subsection{Differences among solitons}

\begin{itemize}
   \item Resonance for solitons composed of spin-0 particles occurs via a dimension-5 operator, where the scalar or pseudo-scalar particle couples with two photons with interacting Lagrangian $\mathcal{L}_{\text{int}} = -(g/4)\phi F_{\mu\nu}F^{\mu\nu}$ and $\mathcal{L}_{\text{int}} = -(g/4)\phi F_{\mu\nu}\tilde{F}^{\mu\nu}$, respectively. By contrast, operators $\Ocal_1$, $\Ocal_2$, and $\Ocal_3$ for the studied cases in tensor solitons and those which lead to resonance for the vector case\,\cite{Amin:2023imi} involve the interaction of two massive spin-$s$ particles with two photons. To be more specific, authors in \cite{Hertzberg:2018zte} analyzed the resonance phenomenon coming from axion condensates. Since the axion couples to electromagnetism via a triangle fermion loop, which connects one axion with two photons, the growth rate of photons is partially proportional to $(g\bar{\phi})$ being peaked at $\approx m/2$. For vector and tensor solitons, the growth rate runs proportional to $(g^2\bar{X}^2)$ and $(g^2\bar{H}^2)$, respectively, being peaked at $\approx m$. Here, it is understood that $\bar{\phi}$ and $\bar{X}$ are the central scalar and vector soliton central amplitudes, respectively.  

   \item If the resonance condition is fulfilled, e.g. $\mu^{(\text{hom})}_{\text{max}} \gtrsim \mu_{\text{esc}}$, scalar solitons would undergo parametric resonance in photons. This occurs because the dimension-5 operators involved allow for an equation of motion for the electromagnetic vector potential in the $k$-space, which exhibits periodic oscillation in the boson field.  Conversely,  one may build dimension-6 operators leading to the interaction between photons and spin-1 or spin-2 particles, which do not lead to resonance. For  circularly-polarized vector soliton, the operators $X_{\sigma}X^{\sigma}F_{\mu\nu}F^{\mu\nu}$ and 
$X_{\sigma}X^{\sigma}F_{\mu\nu}\tilde{F}^{\mu\nu}$ do not give rise 
parametric resonance for any means. For the four possible polarization states in circularly-polarized tensor solitons ($\pm$ 1 and $\pm$ 2), only the third operator among those studied —$H_{\alpha\beta}H^{\beta\mu}F_{\mu\nu}\tilde{F}^{\alpha\nu}$— gives rise resonance for the particular polarization states $\pm 1$. Throughout these instances for both vector and tensor solitons, the lack of resonance is inherently connected to the resultant static nature of the spin-$s$ boson field in the equation of motion for the electromagnetic vector potential in $k$-space.

   \item The emitted radiation from scalar solitons is always unpolarized, meaning it is perfectly isotropic in all directions. In contrast, the electromagnetic radiation emitted by vector and tensor solitons can exhibit a defined polarization. This polarization strictly depends on the original polarization state of the soliton       —whether it is extremely or fractionally polarized— as well as how the spin-$s$ boson particles couple to electromagnetism. For example, circularly and linearly polarized vector solitons may undergo resonance of photons when dark photons couple to photons
   with operators $F_{\mu\sigma}F^{\nu\sigma}X_{\nu}X^{\mu}$ and $\tilde{F}_{\mu\sigma}\tilde{F}^{\nu\sigma}X_{\nu}X^{\mu}$\,\cite{Amin:2023imi}. Similarly, for tensor solitons, we see that an operator $F_{\mu\nu}F^{\alpha\nu}H_{\alpha\beta}H^{\beta\mu}$ may lead to polarized radiation. 

\end{itemize}

\section{Summary and outlook}

In this study, we investigate, for the first time, the phenomenon of parametric resonance in solitons, focusing on a light spin-2 particle as a potential candidate for all or part of dark matter in the Universe. This work is grounded in the framework of effective non-relativistic theory. Our goal is to analyze the key characteristics of this phenomenon, including its potential detectability, as well as the similarities and differences compared to the previously studied parametric resonance in scalar \cite{Hertzberg:2018zte, Hertzberg:2020dbk, Levkov:2020txo, Chung-Jukko:2023cow} and vector solitons \cite{Amin:2023imi}.

Solitons composed of spin-$s$ light boson particles are generally characterized as long-lived, coherent, and spatially localized structures. Due to their unique properties, solitons have the potential to serve as astrophysical laboratories within galaxies. Three main features linked to solitons make them extremely interesting in the context of parametric resonance. Apart from the fact that they are long-lived configurations which could form a significant fraction of dark matter in galactic halos, their density may exceed many orders of magnitude that of the local dark matter neighborhood. For instance, if we consider a light boson particle with a mass of approximately \(m \sim 10^{-6}\, \text{eV}\), solitons with a typical mass of about \(M_{\text{sol}} \sim 10^{-9}\,M_{\odot}\) and a radius of \(R_{\text{sol}} \sim 200\,\text{km}\) exhibit an average density of \(3M_{\text{sol}}/(4\pi R^3_{\text{sol}}) \sim 10^{-8}\,M^4_{\text{sol}} \left(m/m_{\text{pl}}\right)^6\). This density is roughly \(O(10^{26})\) times greater than the local dark matter density \cite{Read:2014qva}. Such a high density would enhance any potential weak interactions between dark matter particles and standard model particles. Furthermore, since solitons are coherently oscillating astrophysical objects, they may exhibit the parametric resonance phenomenon when dark matter particles interact with electromagnetism.

The interactions between a massive dark tensor and photon fields, \(H_{\mu\nu}(x)\) and  \(A_{\mu}(x)\) respectively, present an opportunity to study the electromagnetic radiation emitted by dark tensor solitons in the context of parametric resonance of photons. We examine a direct coupling between spin-2 particles and electromagnetism in the effective field theory framework. This coupling is described by the Lagrangian density \(\Lscr_\mathrm{int} = g^2 \Ocal_i\), where \(\Ocal_i\) represents a defined dimension-6 operator and \(g\) is the coupling constant. Ensuring electromagnetic gauge invariance, we identify three families of operators: 
 \( F_{\mu\nu} F_{\alpha\beta} H_{\sigma\theta} H_{\gamma\epsilon}\)\,,  
\( F_{\mu\nu}  H_{\alpha\beta} H_{\sigma\theta} H_{\gamma\epsilon} H_{\delta\xi}\)\,, and\, 
\( F_{\mu\nu} \partial_{\alpha} \partial_{\beta} H_{\sigma\theta} H_{\gamma\epsilon}\). We do not explore the second and third families, which involve only a single factor of the electromagnetic field \(A_{\mu}\). This is because the power of radiation emitted by operators acting as sources for \(A_{\mu}\) has been previously shown to be exponentially suppressed in the case of scalar solitons \cite{Amin:2021tnq}. 
Although we do not attempt to consider all possible operators within the first family, we focus on a limited set of operators sufficient to capture the key characteristics of the resonance phenomenon. These characteristics include aspects such as frequency, polarization, and spatial patterns of radiation. The dimension-6 operators being considered are as follows:
$\Ocal_1 = -(1/2)F_{\mu\nu}\tilde{F}^{\mu\nu}H_{\alpha\beta}H^{\alpha\beta}\,$,
$\Ocal_2 = -(1/2)F_{\mu\nu}F^{\mu\nu}H_{\alpha\beta}H^{\alpha\beta}\,$,
$\Ocal_3 = F_{\mu\nu}F^{\alpha\nu}H_{\alpha\beta}H^{\beta\mu}\,.$

In the non-relativistic regime for the dark tensor field, these couplings may give rise to a system of equations of motion for the Fourier modes of the electromagnetic field, wherein the dark tensor field induces a periodically oscillating pump that supports the parametric resonance of photons. We conduct analytical and numerical Floquet analyses to assess the feasibility and key characteristics of resonance in dark tensor solitons. When resonance occurs, the Fourier modes of the electromagnetic field grow exponentially, represented as \( | \Avec(t) | \propto e^{\mu_{\text{max}} t} \), where \( \mu_{\text{max}} \) denotes the typical growth rate or maximal real part of the Floquet exponent across all directions of the outgoing radiation. The occurrence of resonance in solitons depends on factors such as the form of the operator that governs the coupling between the tensor field and electromagnetism, the soliton's polarization state, and the coupling strength.

We find that operators $\Ocal_1$ and $\Ocal_2$ do not lead to resonance in circularly polarized tensor solitons (polarization states $\pm 1$ and $\pm 2$). Similarly, the operator $\Ocal_3$ is not linked with resonance in circularly polarized tensor solitons 
with $\pm 2$ polarization states. The reason is that the time dependence of the tensor field is absent for these cases in the equations of motion for the Fourier modes of the electromagnetic field. For instance, this is easy to see for $\Ocal_1 = -(1/2)F_{\mu\nu}\tilde{F}^{\mu\nu}H_{\alpha\beta}H^{\alpha\beta}\,$ in linearly polarized solitons where  $|\Hvec(t)|^2 = \bar{H}^2/2$.

The operators $\Ocal_1$ and $\Ocal_2$ lead to a parametric resonance in linearly polarized tensor solitons. The radiation emitted shows parametrically a maximum growth rate and corresponding bandwidth equal to $\mu_{\text{max}} \sim g^2\bar{H}^2m$ and $\Delta k \sim g^2\bar{H}^2m$, respectively, with an associated band center $k_0 = m + O(m g^4\bar H^4)$. The outgoing radiation is unpolarized for both operators. The operator $\Ocal_3$ leads to \textcolor{black}{parametric resonance} in linearly polarized tensor solitons and circularly polarized solitons with polarization state equal to $\pm 1$. In both cases, the radiation emitted shows parametrically a maximum growth rate and corresponding bandwidth equal to $\mu_{\text{max}} \sim g^2\bar{H}^2m$ and $\Delta k \sim g^2\bar{H}^2m$, respectively, with a corresponding band center $k_0 = m + \mathcal{O}(m g^2\bar H^2)$. However, the outgoing radiation now shows a defined polarization pattern. For linearly polarized solitons, the radiation is mainly linearized, while for circularly polarized solitons (polarization state equal to $\pm 1$), it is primarily circularly
polarized.

We notice that the maximum growth rate, bandwidth, and band center for resonance are parametrically the same for vector\,\cite{Amin:2023imi} and tensor solitons,  
$\mu_{\text{max}} \sim g^2\bar{H}^2m$, $\Delta k \sim g^2\bar{H}^2m$, and $k_0 \sim m $, but different for scalar solitons,  $\mu_{\text{max}} \sim g\bar{\phi}m$, $\Delta k \sim g\bar{\phi}m$, $k_0 \sim m/2 $\,\cite{Hertzberg:2018zte}. The reason comes from the interactions between dark matter particles and electromagnetism. While for scalar particles, the family of operators of the form $\phi F_{\mu\nu} F_{\alpha\beta}$ couples one spin-0 field with two photon fields, the family of operators for vector, $F_{\mu\nu} F_{\alpha\beta} X_{\sigma}X_{\theta}$\,\cite{Amin:2023imi}, and tensor solitons, $ F_{\mu\nu} F_{\alpha\beta} H_{\sigma\theta}H_{\gamma\epsilon}$, couple two spin-$s$ fields with two-photon fields. There are more families of operators available for vector and tensor solitons, but for them, the radiation emitted is $O(g^4)$ or is exponentially suppressed by a factor $mR_{\text{sol}} \gg 1$.  

Suppose the interaction between the dark tensor field and the photon field, along with the soliton polarization state, enables resonance. In that case, it is found that tensor solitons will resonate if the strength of the coupling exceeds a critical value, denoted as \( g_{\text{crit}} \). Alternatively, this can also be expressed in terms of mass; resonance occurs when the total mass of the soliton exceeds a critical value \( M_{\text{sol, crit}} \). These equivalent resonance conditions can be written as $ g_{\text{crit}} \approx \left( \textcolor{black}{\bar\mu}/m \right)^{-3/4}m_{\text{pl}}^{-1}\,$, and  $M_{\text{sol,crit}} \approx 9 \times 10^{-10} M_{\odot} \left(10^{-6}\text{eV}/m\right) \left(10^{-10}\text{GeV}^{-1}/g_{\text{crit}}\right)^{2/3}\,,$ where $(\textcolor{black}{\bar\mu}/m) \ll 1$ is the characteristic ratio between the chemical potential and the spin-2 particle mass. Similar behavior for resonance was previously observed for scalar\,\cite{Hertzberg:2018zte, Levkov:2020txo} and vector solitons\,\cite{Amin:2023imi}. 

 The resonance condition establishes an upper bound for the tensor soliton mass today within galactic halos among those that can experience resonance after formation. Since multiple dimension 6-operator could lead to resonance simultaneously and polarization states of solitons should be approximately equally distributed for energetic reasons (on the absence of self-interactions, soliton energies for different states of polarization are degenerate \cite{Jain:2021pnk}), we should anticipate a significant fraction of solitons with masses just below the threshold in galactic halos. However, this is not the end of the story, as mergers between solitons in galactic halos could provide a way to generate super-critical mass solitons.  By carefully examining the velocity distributions of solitons in the Galaxy and the necessary conditions for effective mergers following collisions, we estimate the Galactic merger rate to be at most about $6 \times 10^{-8}\,\text{year}^{-1}$. This low rate is primarily because typical solitons carry too much energy for a successful merger.
 Since the merger rate is highly dependent on galactic dark matter halo profiles, $\Gamma_{\text{merg}} \propto \rho^2_{\text{halo}}$, it is compelling to recalculate this rate considering the presence of spiky\,\cite{Gondolo:1999ef, Bertone:2002je} dark matter halo profiles in galaxies triggered by the presence of supermassive black holes at their
centers\,\cite{Kormendy:1995er, Kormendy:2013dxa}, as discussed in a different context by  Ref.\,\cite{Amaral:2023ekd}. We will revisit this point a few paragraphs below. 

After a soliton merger, let us consider that the resultant soliton holds a supercritical mass, leading to a parametric resonance of photons. In the scenario of an instantaneous burst after the configuration settles down to a ground state soliton, the time scale is estimated as the inverse of the maximum growth rate, e.g. $ \tau_{\text{inst}}  \sim 120\,\mu\text{s} (10^{-6}\,\text{eV}/m)
(10^{-10}\,\text{GeV}^{-1}/g_{\text{crit}})^{-2/3}$. In the soliton rest frame, the radiation emitted is highly monochromatic and peaks within the radio band with a central frequency $\nu_{0} \sim 240\,\text{MHz} (10^{-6}\,\text{eV}/m)^{-1}$ and bandwidth $\Delta\nu_{0}\sim 3\,\text{kHz} (10^{-6}\,\text{eV}/m)^{-1}(10^{-10}\,\text{GeV}^{-1}/g_{\text{crit}})^{2/3}$. \textcolor{black}{Considering that the radiated mass is approximately \(O(M_{\text{sol, crit}})\), and taking into account the cosmic redshift of the signal's wavelength due to cosmic expansion, along with a coupling constant value of \(g \sim g_{\text{crit}} = 10^{-10}\,\text{GeV}^{-1}\), the predicted spectral flux density associated with the signal is expected to be several orders of magnitude higher than the typical sensitivity of current and future radio telescopes\,\cite{SKA1}. This is illustrated in Eq.\,(\ref{Eq:UpperSB})}. In simpler terms, if an extragalactic or galactic tensor soliton were to explode in electromagnetic bursts, the signal strength would not pose a concern from a detection standpoint. Furthermore, in principle, the polarization of the emitted radiation could also be detected by ground-based facilities, providing a compelling method to distinguish the radiation source. This is significant because scalar solitons are connected to unpolarized radiation only through the parametric resonance of photons. A detailed analysis of the feasibility of this compelling possibility is left for future work.    

\textcolor{black}{The validity of our EFT requires that $g^2\bar{H}^2\ll 1$, Eq.\,(\ref{eq:EFTcond}), which can be seen as an upper bound for the coupling constant. Via Eqs.\,(\ref{eq:Hprofile}), this condition takes the simple form $gm_{\text{pl}}\ll (\bar{\mu}/m)^{-1}$, right side on Eq.\,(\ref{Rcond}). Assuming the core-halo relation given in Sec. 2.2, we consider nucleation of tensor solitons at $100 \lesssim z \lesssim 10$ in dark matter minihalos with a mass range of   
$10^{-4}\,M_{\odot} \lesssim M_{\text{Mh}} \lesssim 10^{8}\,M_{\odot}$. For a spin-2 particle mass that ranges as
$10^{-7}\,\text{eV} \lesssim m \lesssim 10^{-3}\,\text{eV}$,
the tensor soliton masses spread as 
$10^{-14}\,M_{\odot} \lesssim M_{\text{sol}} \lesssim 10^{-5}\,M_{\odot}$. Equations\, (\ref{eq:Msol}) and \,(\ref{Rcond}) allow us to estimate an upper bound for the coupling.} 

\textcolor{black}{In general, the core-halo relation tells us that for a fixed redshift and minihalo mass, the nucleated soliton mass increases as the mass of the spin-2 particles decreases. Both tendencies tend to compensate in Eq.\,(\ref{eq:Msol}), and the coupling upper bound results in being weakly dependent on the mass of the spin-2 particles. For example, take $z=100$ and $ M_{\text{Mh}} = 10^{-4}\,M_{\odot}$. We have $g  \ll 10^{-5}\,\text{GeV}^{-1}$ for $m \sim 10^{-3}\,\text{eV}$ and  $g  \ll 10^{-4}\,\text{GeV}^{-1}$ for $m \sim 10^{-7}\,\text{eV}$.}

\textcolor{black}{We are currently unaware of any phenomenological constraints regarding the coupling of the proposed dimension-6 operators\,\footnote{\textcolor{black}{Our dimension-6 operators working in the context of EFT, suggested in Ref. \cite{Jain:2021pnk}, were chosen primarily to compare with previous results on vector solitons\,\cite{Amin:2023imi}. Here we mention that the predicted coupling $\alpha/m_{\text{pl}}$, where $\alpha$ is a dimensionless quantity, between spin-2 particles and electromagnetism in ghost-free bigravity takes the effective form $E_iE_j \pm B_iB_j$ and $E_iB_j \pm B_iE_j$ \,\cite{Han:1998sg}. The $\alpha/m_{\text{pl}}$ coupling in this theory has been constrained for fifth force searches and tests of the equivalence principle\,\cite{Marzola:2017lbt}. It could be interesting to extend our analysis for the mentioned coupling in the framework of ghost-free bigravity. We leave this task for future work.}}; however, these constraints could be determined by considering specific astrophysical scenarios. One potential scenario involves observing radio emissions from the merger of tensor solitons within spiky dark matter halos surrounding supermassive black holes. It is well established that almost all large galaxies harbor supermassive black holes at their centers \cite{Kormendy:1995er, Kormendy:2013dxa}. Under certain conditions, the growth of these astrophysical entities can lead to a highly concentrated dark matter profile in the central regions of galaxies \cite{Gondolo:1999ef, Bertone:2002je, Zhang:2025mdl}. In such regions, the increased dark matter density can result in significant merger events, during which solitons may exceed a critical mass and emit radiation through parametric resonance. This process produces brief, narrowband, and highly energetic bursts of radiofrequency radiation, signals that fast radio burst experiments could detect. If corresponding radio transients are not observed in fast radio burst surveys, this would place constraints on the fraction of solitonic dark matter or the coupling strength constant. This idea has already been explored in the context of vector solitons\,\cite{Dorian:2025}.}

Since our analysis is based on the instantaneous burst assumption, it is necessary to improve this approach to obtain more accurate signal predictions. We are exploring configurations where the typical soliton's light-crossing time, which corresponds to the order of the soliton's length scale, is significantly longer than the typical resonance time scale. This feature suggests that radiation may leak from tensor solitons through weaker bursts as the configuration stabilizes into the final ground-state condensate. To address this issue, numerical simulations that consider the backreaction of the radiative process in the soliton configuration will be beneficial, similar to what was done in Ref.\,\cite{Amin:2021tnq} regarding dipole radiation in scalar solitons.

To finalize, we would like to emphasize that our research has revealed that dark tensor solitons, which are composed of light spin-2 particles, can undergo parametric resonance of photons. Similar to vector solitons\,\cite{Amin:2023imi}, we have demonstrated that the emitted radiation can exhibit a defined polarization pattern that encodes both the soliton's polarization state and the nature of the interaction between the spin-2 field and photons. This feature is not observed in the case of scalar solitons\,\cite{Hertzberg:2018zte, Levkov:2020txo}, creating a compelling opportunity for detecting the underlying spin of dark matter.

\begin{acknowledgments}

E.D.S. expresses gratitude to Dorian W.P. Amaral from the Department of Physics and Astronomy at Rice University, USA, for providing the numerical results from Monte Carlo simulations needed to determine the relative velocity distribution functions in Sec.\,\ref{Sec:mergerrate}. E.D.S. would like to extend a special thanks to Hernán Gonzalez from the Theoretical Physics Group at Universidad San Sebastián, Campus Ciudad Empresarial, Chile, and Andrew Long from the Department of Physics and Astronomy at Rice University, USA, for enlightening discussions.  E.D.S.
acknowledges support from the FONDECYT project 1251141 (Agencia Nacional de Investigaci\'on y Desarrollo, Chile).
 \end{acknowledgments}

\appendix
\label{Appendix}

\section{General Floquet analysis}
\label{app:general}

After imposing the Coulomb constraint $\kvec\cdot\Avec=0$ on Eq.\,(\ref{eq:OPQ_equation}), the obtained reduced system can be written as
\begin{equation}
\ddot{\Avec} + \tilde{\mathbb{O}}^{-1}\tilde{\mathbb{P}}\dot{\Avec} + \tilde{\mathbb{O}}^{-1}\tilde{\mathbb{Q}}\Avec = 0\,,
\label{eq:reduced}
\end{equation}
where $\tilde{\mathbb{O}}^{-1}\tilde{\mathbb{O}}=\mathds{1}$ is understood. The fact that matrices $\tilde{\mathbb{O}}^{-1}\tilde{\mathbb{P}}$ and $\tilde{\mathbb{O}}^{-1}\tilde{\mathbb{Q}}$ correspond to periodic functions with period $T = 2\pi/\omega_0$ allow us 
 to expand them as a Fourier series $\tilde{\mathbb{O}}^{-1}\tilde{\mathbb{P}} = \sum_{l'}[\tilde{\mathbb{O}}^{-1}\tilde{\mathbb{P}}]_{l'} e^{il'\omega_0t}$ and $\tilde{\mathbb{O}}^{-1}\tilde{\mathbb{Q}} = \sum_{l'}[\tilde{\mathbb{O}}^{-1}\tilde{\mathbb{Q}}]_{l'} e^{il'\omega_0t}$. Similarly, the vector potential in the $k$-space may also be expanded as $\Avec = \sum_{l}{\tilde\Avec}_{l}e^{il\omega_0t}$, where ${\tilde\Avec}_{l}$ is a slowly varying function in time, i.e. $\ddot{\tilde\Avec}_{l} \approx 0$.
Here $l$ and $l'$ are integers.
Under all these replacements, Eq.\,(\ref{eq:reduced}) becomes
\begin{equation}
\sum_l \left( 2il\omega_0\dot{\tilde\Avec}_l -(l\omega_0)^2{\tilde\Avec}_l\right)e^{il\omega_0t} + \sum_l \sum_{l'} [\tilde{\mathbb{O}}^{-1}\tilde{\mathbb{P}}]_{l'}
\left( \dot{\tilde\Avec}_l + il\omega_0{\tilde\Avec}_l \right)e^{i(l'+l)\omega_0t} + \sum_l \sum_{l'} [\tilde{\mathbb{O}}^{-1}\tilde{\mathbb{Q}}]_{l'}{\tilde\Avec}_l e^{i(l+l')\omega_0t}=0\,.
\end{equation}
Multiplying the whole equation by $e^{-i\omega_0t}$ and integrating from $0$ to $2\pi$, we have

\begin{align}
&\left( 2il\omega_0\mathds{1}+[\tilde{\mathbb{O}}^{-1}\tilde{\mathbb{P}}]_0  \right) \dot{\tilde\Avec}_l + \left( [\tilde{\mathbb{O}}^{-1}\tilde{\mathbb{Q}}]_0 + il\omega_0[\tilde{\mathbb{O}}^{-1}\tilde{\mathbb{P}}]_0 - (l\omega_0)^2\mathds{1}\right){\tilde\Avec}_l\nonumber\\
&\hspace{2cm}+ \sum_{l'\neq 0} [\tilde{\mathbb{O}}^{-1}\tilde{\mathbb{P}}]_{l'} \dot{\tilde\Avec}_{l-l'} + \sum_{l'\neq 0} \left([\tilde{\mathbb{O}}^{-1}\tilde{\mathbb{Q}}]_{l'}+i[\tilde{\mathbb{O}}^{-1}\tilde{\mathbb{P}}]_{l'}(l-l')\omega_0  \right) {\tilde\Avec}_{l-l'} = 0\,,
\end{align}
where we have assumed that ${\tilde\Avec}_l$, $\tilde{\mathbb{O}}$, $\tilde{\mathbb{P}}$, and $\tilde{\mathbb{Q}}$ are slowing varying.  
Since the lowest frequencies are dominant in the first instability band, we take $(l'=+2, l=+1)$ and $(l'=-2,l=-1)$ in the above equation. Dropping all higher harmonics, we have the following coupled pair of differential equations for the lowest frequency modes:
\begin{equation}
\begin{pmatrix}
    \dot{\tilde{A}}_{\omega_0} \\
    \dot{\tilde{A}}_{-\omega_0}
\end{pmatrix}
 = \tilde{\mathbb{M}}
    \begin{pmatrix}
         \tilde{A}_{\omega_0}\\
         \tilde{A}_{-\omega_0}
    \end{pmatrix}
    \;,
\end{equation}
where $\tilde{\mathbb{M}} = \tilde{\mathbb{M}}_a^{-1}\tilde{\mathbb{M}}_b$ and 
\begin{align}
\tilde{\mathbb{M}}_a^{-1}&=
 \begin{pmatrix}
         2i\omega_0\mathds{1}+[\tilde{\mathbb{O}}^{-1}\tilde{\mathbb{P}}]_0  &[\tilde{\mathbb{O}}^{-1}\tilde{\mathbb{P}}]_2\\
         [\tilde{\mathbb{O}}^{-1}\tilde{\mathbb{P}}]_{-2}  &-2i\omega_0\mathds{1}+[\tilde{\mathbb{O}}^{-1}\tilde{\mathbb{P}}]_0
    \end{pmatrix}^{-1} 
    \;,\label{eq:Ma}\\
    \tilde{\mathbb{M}}_b&=
 \begin{pmatrix}
         -[\tilde{\mathbb{O}}^{-1}\tilde{\mathbb{Q}}]_0 -i\omega_0[\tilde{\mathbb{O}}^{-1}\tilde{\mathbb{P}}]_0  + \omega_0^2\mathds{1} &-[\tilde{\mathbb{O}}^{-1}\tilde{\mathbb{Q}}]_2 + i\omega_0[\tilde{\mathbb{O}}^{-1}\tilde{\mathbb{P}}]_{2}\\
         -[\tilde{\mathbb{O}}^{-1}\tilde{\mathbb{Q}}]_{-2} -i\omega_0[\tilde{\mathbb{O}}^{-1}\tilde{\mathbb{P}}]_{-2} &-[\tilde{\mathbb{O}}^{-1}\tilde{\mathbb{Q}}]_0 + i\omega_0[\tilde{\mathbb{O}}^{-1}\tilde{\mathbb{P}}]_0  + \omega_0^2\mathds{1}
    \end{pmatrix} \label{eq:Mb}
    \;.
\end{align}
We explore now the structure of $\tilde{\mathbb{M}}$. Suppose that $\tilde{\mathbb{M}}^{-1}_a \equiv (\mathbb{A} + g^2\mathbb{B})^{-1} = \mathbb{A}^{-1} + g^2\mathbb{S}$, where $\mathbb{A}, \mathbb{B}, \mathbb{S}$ are matrices to be determined. Since $(\mathbb{A}^{-1}+g^2\mathbb{S})(\mathbb{A}+g^2\mathbb{B}) = \mathds{1} + g^2\mathbb{A}^{-1}\mathbb{B} + g^2\mathbb{S}\mathbb{A} + \mathcal{O}(g^4) = \mathds{1}$, we have $\mathbb{S} = -\mathbb{A}^{-1}\mathbb{B}\mathbb{A}^{-1}$ to second order in the coupling. Comparing with Eq.\,(\ref{eq:Ma}), we have 
\begin{equation}
\tilde{\mathbb{M}}_a^{-1} = \mathbb{A}^{-1}-g^2\mathbb{A}^{-1}\mathbb{B}\mathbb{A}^{-1}\,\,\,\,\,\text{with}\,\,\,\,\, 
\mathbb{A}^{-1} = 
 \begin{pmatrix}
         -\frac{i}{2\omega_0}\mathds{1} &0\\
         0 & \frac{i}{2\omega_0}\mathds{1}
    \end{pmatrix} 
    \,\,\,\text{and}\,\,\, 
    \mathbb{B} = \frac{1}{g^2}
 \begin{pmatrix}
         [\tilde{\mathbb{O}}^{-1}\tilde{\mathbb{P}}]_0 &[\tilde{\mathbb{O}}^{-1}\tilde{\mathbb{P}}]_2\\
         [\tilde{\mathbb{O}}^{-1}\tilde{\mathbb{P}}]_{-2} & [\tilde{\mathbb{O}}^{-1}\tilde{\mathbb{P}}]_0
    \end{pmatrix}\,, 
\end{equation}
where $[\tilde{\mathbb{O}}^{-1}\tilde{\mathbb{P}}]_0 = \mathcal{O}(g^2)$. Following the same idea, we write 
now $\tilde{\mathbb{M}}_b= \mathbb{C}+g^2\mathbb{D}$, where $\mathbb{C}, \mathbb{D}$ are matrices to be determined. Comparing with Eq.\,(\ref{eq:Mb}), we have 
\begin{equation}
\mathbb{C}=
\begin{pmatrix}
         (\omega_0^2-k^2)\mathds{1} &0\\
         0 & (\omega_0^2-k^2)\mathds{1}
    \end{pmatrix} 
    \,\,\,\text{and}\,\,\, 
\mathbb{D} = \frac{1}{g^2}
 \begin{pmatrix}
         -i\omega_0[\tilde{\mathbb{O}}^{-1}\tilde{\mathbb{P}}]_0 - \overline{[\mathbb{O}^{-1}\mathbb{Q}]}_0 &i\omega_0[\tilde{\mathbb{O}}^{-1}\tilde{\mathbb{P}}]_2 - [\tilde{\mathbb{O}}^{-1}\tilde{\mathbb{Q}}]_2\\
         -i\omega_0[\tilde{\mathbb{O}}^{-1}\tilde{\mathbb{P}}]_{-2} - [\tilde{\mathbb{O}}^{-1}\tilde{\mathbb{Q}}]_{-2} & i\omega_0[\tilde{\mathbb{O}}^{-1}\tilde{\mathbb{P}}]_0 - \overline{[\mathbb{O}^{-1}\mathbb{Q}]}_0
    \end{pmatrix}\,, 
\end{equation}
where $\overline{[\tilde{\mathbb{O}}^{-1}\tilde{\mathbb{Q}}]}_0=[\tilde{\mathbb{O}}^{-1}\tilde{\mathbb{Q}}]_0-k^2\mathds{1}=\mathcal{O}(g^2)$. Since $\mathbb{A}^{-1}, \mathbb{B}, \mathbb{C}, \mathbb{D} $ matrices are now known, we are ready to calculate $\tilde{\mathbb{M}}$ to the second order in the coupling. Noting that $\tilde{\mathbb{M}} = (\mathbb{A}^{-1}-g^2\mathbb{A}^{-1}\mathbb{B}\mathbb{A}^{-1})(\mathbb{C}+g^2\mathbb{D}) = \mathbb{A}^{-1}\mathbb{C}+g^2(\mathbb{A}^{-1}\mathbb{D}-\mathbb{A}^{-1}\mathbb{B}\mathbb{A}^{-1}\mathbb{C})+\mathcal{O}(g^4)$, we find
\begin{equation}
\tilde{\mathbb{M}} = 
\begin{pmatrix}
        -\frac{i(\omega_0^2-k^2)}{2\omega^2_0}\mathds{1} - \frac{(\omega_0^2+k^2)}{4\omega_0^2}[\tilde{\mathbb{O}}^{-1}\tilde{\mathbb{P}}]_0 + \frac{i}{2\omega_0}[\overline{\mathbb{O}^{-1}\mathbb{Q}}]_0 & 
        \frac{(\omega_0^2+k^2)}{4\omega^2_0}[\tilde{\mathbb{O}}^{-1}\tilde{\mathbb{P}}]_2 + 
        \frac{i}{2\omega_0}[\tilde{\mathbb{O}}^{-1}\tilde{\mathbb{Q}}]_2
        \\
        \frac{(\omega_0^2+k^2)}{4\omega^2_0}[\tilde{\mathbb{O}}^{-1}\tilde{\mathbb{P}}]_{-2} - 
        \frac{i}{2\omega_0}[\tilde{\mathbb{O}}^{-1}\tilde{\mathbb{Q}}]_{-2}
         & 
          \frac{i(\omega_0^2-k^2)}{2\omega^2_0}\mathds{1} - \frac{(\omega_0^2+k^2)}{4\omega_0^2}[\tilde{\mathbb{O}}^{-1}\tilde{\mathbb{P}}]_0 - \frac{i}{2\omega_0}[\overline{\mathbb{O}^{-1}\mathbb{Q}}]_0
    \end{pmatrix}\,, 
\end{equation}
where matrix entries are $\mathcal{O}(g^2)$ and the matrix eigenvalues are given by the two pairs of Floquet exponents, e.g., $\mu_s$ with $s={1,2,3,4}$. This result agrees with that shown in Ref.\,\cite{Amin:2023imi}, but where most intermediate steps were omitted. 

\section{Detailed Floquet analysis of a few examples}
\label{app:details}

While the general methodology for obtaining the corresponding Floquet solutions is outlined in the main text, we provide additional details on some interesting cases from a computational perspective.

\subsection{Homogeneous dark tensor field  ${{\bf{\textit{H}}}^{(0)}}$ for $\mathcal{O}_2$}
\label{apphomo:H002}

The equation of motion for the vector potential in the $k$-space at order $O(g\bar{H})^2$ for a homogeneous dark tensor field with null polarization state is given by Eq.\,(\ref{eq:OPQ_equation}):
\begin{equation}
\ddot{\Avec} - 2g^2\bar{H}^2 m\, \text{sin}(2 m t)\dot{\Avec} + k^2\Avec = 0\,.
\label{eq:H002}
\end{equation}
We express the trigonometric function using exponential terms and perform a harmonic expansion of $A \equiv  A_i(t)$  as follows:
\begin{equation}
    A(t) = \sum_{l=-\infty}^{\infty} \tilde{A}_{l}(t) \, e^{i l m t} 
    \;, 
\label{eq:Aharmonic}
\end{equation}
where  $\tilde{A}_{l}(t)$ is a slowly varying function in time, i.e.  
$\ddot{\tilde{A}}_{l}\approx 0$. In the $g^2\bar{H}^2\ll 1$ regime, there exists a spectrum of narrow resonance bands equally spaced at  $k^2\approx n^2m^2$ for $n=1,2,3, \ldots$. We recollect terms proportional to 
$e^{i l m t}$, $e^{i(l + 2)  m t}$, and $e^{i (l - 2 ) m t}$. In each sum, we proceed to change the summation index to make all these terms proportional to
$e^{i l m t}$. We integrate over time within the interval $0 \leq t \leq 2\pi/m$. Since the resonance phenomenon is dominated by the first instability band, we evaluate the obtained equation at $l = \pm 1$. As a result,
the resultant system of differential equations to be solved is given by 
\begin{equation}
\begin{pmatrix}
    \dot{\tilde{A}}_{+} \\
    \dot{\tilde{A}}_{-}
\end{pmatrix}
    = 
    \begin{pmatrix}
         \tilde{\mathbb{M}}_{11}  &\tilde{\mathbb{M}}_{12}\\
         -\tilde{\mathbb{M}}_{12}  &-\tilde{\mathbb{M}}_{11}
    \end{pmatrix} 
    \begin{pmatrix}
         \tilde{A}_{+}\\
         \tilde{A}_{-}
    \end{pmatrix}
    \;,
\end{equation}
where
\begin{align}
\tilde{\mathbb{M}}_{11} &= \frac{i}{2m}\left(k^2-m^2  \right)\,,\\
\tilde{\mathbb{M}}_{12} &= -\frac{i}{2m}\left( g^2\bar{H}^2m^2 + \frac{1}{2} g^2\bar{H}^2  (k^2 - m^2)\right)\,.
\end{align}
The eigenvalues of the above matrix correspond to the Floquet exponents. There are two eigenvalues with identical magnitude but different signs. The positive eigenvalue reads as
\begin{align}
  \mu_{\kvec,\text{max}} =  \sqrt{-\frac{(k^2-m^2)^2}{4m^2}+\frac{g^4\bar{H}^4(k^2+m^2)^2}{16m^2} }\,.
  \label{eq:H002mumax}
  \end{align}
The edges of the first instability are placed at   $\mu_{\kvec,\text{max}}(k_{l,{\rm edge}})=\mu_{\kvec,\text{max}}(k_{r,{\rm edge}})=0$, where $k_{l,{\rm edge}}$ and $k_{r,{\rm edge}}$ are the left and right edge, respectively, in the $k$-space. From Eq.\,(\ref{eq:H002mumax}), we have
\begin{align}
k_{l,{\rm edge}} &= m \frac{\sqrt{2-g^2\bar{H}^2}}{\sqrt{2+g^2\bar{H}^2}} = m - \frac{1}{2} g^2 \bar{H}^2 m +O(g^4\bar{H}^4m) \,,\\
k_{r,{\rm edge}} &= m \frac{\sqrt{2+g^2\bar{H}^2}}{\sqrt{2-g^2\bar{H}^2}} = m + \frac{1}{2} g^2 \bar{H}^2 m +O(g^4\bar{H}^4m) \,.
\end{align}
The center of the first instability band  $k_0$ and its bandwidth $\Delta k$ are readily calculated using the above equations as
\begin{align}
k_0 &= \frac{(k_{l,{\rm edge}}+k_{r,{\rm edge}})}{2} = m + O(g^4\bar{H}^4m)\,,\\
\Delta k &= (k_{r,{\rm edge}}-k_{l,{\rm edge}}) = g^2 \bar{H}^2 m + O(g^6\bar{H}^6m)\,.
\end{align}
The largest Floquet exponent along all possible wavenumbers is calculated by maximizing Eq.\,(\ref{eq:H002mumax}) in terms of 
$k$. We find,
\begin{equation}
\mu_{\text{max}} = \frac{g^2\bar{H}^2m}{2} + O(g^4\bar{H^4}m)\,.
\end{equation}
The analytical results are in complete agreement with those obtained by numerical calculations.

\subsection{Homogeneous dark tensor field  ${{\bf{\textit{H}}}^{(0)}}$ for $\mathcal{O}_3$}
\label{apphomo:H003}

\begin{figure}[t]
\centering
      \includegraphics[scale=0.5]{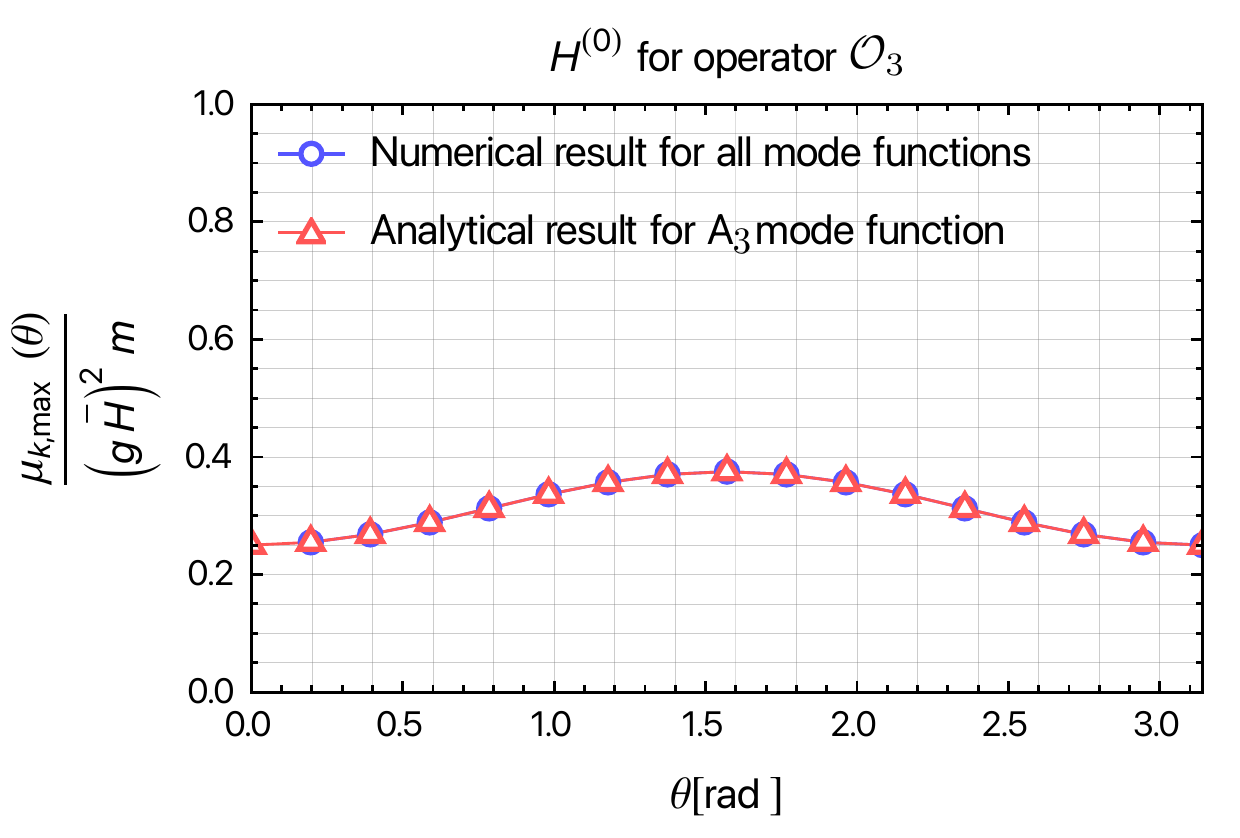}
\caption{\label{ProofRes}
Plot for  $\mu_{\kvec,\text{max}}(\theta)$ in terms of the angle between the wavevector $\kvec$ of the outgoing radiation and the $\hat{\bf z}$ direction, blue circles. We have added the analytical result for $A_3$ electromagnetic Fourier mode, red triangle.
}
\label{Fig:H0O3}
\end{figure}

 After applying the Coulomb gauge condition and using $k_3=k\text{cos}(\theta)$, the equation of motion for $A_3$ decouples from $A_1$ and $A_2$ taking the form
\ba{\label{eq:A3_H0_03}
    &\ddot{A}_3 + \left( \frac{4}{3} g^2 m \bar{H}^2 \text{sin}(2 m t) -   g^2 m \bar{H}^2 \text{cos}^2(\theta)\text{sin}(2 m t) \right)\dot{A}_3\nonumber\\ 
    &+ \left(  k^2 - \frac{1}{6} g^2 k^2 \bar{H}^2   -\frac{1}{2} g^2 k^2 \text{cos}^2(\theta) \bar{H}^2 -\frac{1}{6} g^2 k^2 \bar{H}^2 \text{cos}(2 m t) -\frac{1}{2} g^2 k^2\text{cos}^2(\theta) \bar{H}^2 \text{cos}(2 m t)
    \right)A_3=0\;,
}
where we have kept terms until $O(g\bar{H})^2$.
As explained in the main text, we express the trigonometric functions in terms of exponentials and perform a harmonic expansion of $A_3(t)$ as follows:
\begin{equation}
    A_3(t) = \sum_{l=-\infty}^{\infty} \tilde{A}_{3,l}(t) \, e^{i l m t} 
    \;. 
\label{eq:A3harmonic}
\end{equation}
In this context, $\tilde{A}_{3,l}(t)$ represents a slowly varying function over time, which implies that $\ddot{\tilde{A}}_{3,l} \approx 0$. Here, $l$ is an integer. In the regime where $g^2 \bar{H}^2 \ll 1$, we conduct a harmonic expansion of the electromagnetic modes, similar to what was done in the previous subsection, to obtain
\begin{equation}
\begin{pmatrix}
    \dot{\tilde{A}}_{3,+} \\
    \dot{\tilde{A}}_{3,-}
\end{pmatrix}
    = 
    \begin{pmatrix}
         \tilde{\mathbb{M}}_{11}  &\tilde{\mathbb{M}}_{12}\\
         -\tilde{\mathbb{M}}_{12}  &-\tilde{\mathbb{M}}_{11}
    \end{pmatrix} 
    \begin{pmatrix}
         \tilde{A}_{3,+}\\
         \tilde{A}_{3,-}
    \end{pmatrix}
    \;,
\end{equation}
where
\begin{align}
\tilde{\mathbb{M}}_{11} &= \frac{-i}{288 m}\left( -144k^2 + 144m^2 +24g^2\bar{H}^2k^2 + 72g^2\bar{H}^2k^2\text{cos}^2(\theta)  \right)\,,\\
\tilde{\mathbb{M}}_{12} &= \frac{ig^2\bar{H}^2}{144m} \left( -30k^2 -24m^2 +4g^2\bar{H}^2k^2+18m^2\text{cos}^2(\theta)+9g^2\bar{H}^2k^2\text{cos}^2(\theta)-9g^2\bar{H}^2k^2\text{cos}^4(\theta) \right)\,.
\end{align}
 The eigenvalues of the above matrix give the Floquet exponents linked to the $A_3$ mode function. There is a pair of eigenvalues with identical magnitude but different signs. The positive eigenvalue reads as
\begin{align}
  \mu_{\kvec,\text{max}} =& \frac{1}{1152 \sqrt{2}m} \left[ -( 663552 k^4 - 1327104 k^2 m^2 + 663552 m^4 - 
 552960 g^2  \bar{H}^2 k^4+ 552960 g^2 \bar{H}^2k^2 m^2\right.\nonumber\\
 &\left.+ 
 20736 g^4\bar{H}^4 k^4  - 115200 g^4 \bar{H}^4k^2 m^2  - 
 33984 g^4\bar{H}^4 m^4  - 331776 g^2\bar{H}^2 k^4  \text{cos}(2 \theta)\right.\nonumber\\
 &\left.+ 
 331776 g^2 \bar{H}^2k^2 m^2 \text{cos}(2 \theta) + 
 138240 g^4\bar{H}^4  k^4  \text{cos}(2 \theta) + 
 69120 g^4\bar{H}^4  k^2 m^2  \text{cos}(2 \theta)\right.\nonumber\\
 &\left.+ 
 34560 g^4 \bar{H}^4 m^4  \text{cos}(2 \theta) + 
 20736 g^4\bar{H}^4  k^4  \text{cos}(4 \theta) - 
 5184 g^4\bar{H}^4  m^4 \text{cos}(4 \theta) + O(g\bar{H})^6 )    \right]^{1/2}\,.\label{eq:H003mumax}
\end{align}
The left and right edges of the first instability are calculated from Eq.\,(\ref{eq:H003mumax}) as
\begin{align}
k_{l,{\rm edge}} &= m \frac{\sqrt{24-5g^2\bar{H}^2 + 3g^2\bar{H}^2\text{cos}(2\theta)}}{\sqrt{24-6g^2\bar{H}^2\text{cos}(2\theta)}} = m - \frac{5}{48} g^2 \bar{H}^2 m + \frac{3}{16} g^2 \bar{H}^2 m \text{cos}(2\theta) + O(g^4\bar{H}^4m) \,,\\
k_{r,{\rm edge}} &=m \frac{\sqrt{24+5g^2\bar{H}^2 - 3g^2\bar{H}^2\text{cos}(2\theta)}}{\sqrt{24-20g^2\bar{H}^2-6g^2\bar{H}^2\text{cos}(2\theta)}} = m + \frac{25}{48} g^2 \bar{H}^2 m + \frac{1}{16} g^2 \bar{H}^2 m \text{cos}(2\theta) + O(g^4\bar{H}^4m) \,.
\end{align}
The center of the first instability band  and its bandwidth read as
\begin{align}
k_0 &= \frac{(k_{l,{\rm edge}}+k_{r,{\rm edge}})}{2} = m +  g^2 \bar{H}^2 m \left(\frac{5}{24}+\frac{\text{cos}(2\theta)}{8}\right)  
+ O(g^4\bar{H}^4m)\,,\\
\Delta k &= (k_{r,{\rm edge}}-k_{l,{\rm edge}}) = g^2 \bar{H}^2 m\left(\frac{5}{8}-\frac{\text{cos}(2\theta)}{8} \right) + O(g^4\bar{H}^4m)\,.
\end{align}
Replacing $k_0$ into Eq.\,(\ref{eq:H003mumax}), we estimate the maximum Floquet exponent in terms of the polar angle between the dark tensor polarization and the propagation direction of photons as 
\begin{equation}
  \mu_{\kvec,\text{max}}(\theta) = g^2 \bar{H}^2 m\frac{\sqrt{51/2}}{16} \left(1 - \frac{20\text{cos}(2\theta)}{51}+\frac{\text{cos}(4\theta)}{51}  \right)^{1/2} + O(g^4 \bar{H}^4m)\,.
\end{equation}
As shown in Fig.\,\ref{Fig:H0O3}, this expression corresponds to the largest Floquet exponent across all possible wave vectors. This holds even though $\mu_{\kvec,\text{max}}(\theta)$ was determined using only the electromagnetic field equation of motion for $A_3$.

\section{Floquet analysis for spherical tensor soliton profile}
\label{app:details2}

When considering localized soliton condensates, transitioning to $k$-space for the equation of the vector potential, as we did previously in Eq.~(\ref{eq:OPQ_equation}) for a homogeneous condensate, is not highly recommended. This approach leads to a convolution term that couples the $k$-modes of the field with those of the Fourier electromagnetic modes. A more effective method for addressing the internal structure of tensor solitons is to perform a vector spherical harmonic decomposition of the vector potential $\mathbf{A}(t, \mathbf{x})$. Similar techniques have been applied to oscillons, axion solitons, and vector solitons in Refs.~\cite{Hertzberg:2010yz}, ~\cite{Hertzberg:2018zte}, and ~\cite{Amin:2023imi}, respectively. 

As an example, we utilize the operator $\mathcal{O}_2 = -(1/2) F_{\mu\nu} F^{\mu\nu} H_{\alpha\beta} H^{\alpha\beta}$, which involves a contraction of indices between the covariant and contravariant forms of the electromagnetic tensor. The coordinate space representation of Eq.\,(\ref{eq:OPQ_equation}) reads as
\begin{equation}
\begin{adjustbox}{max width=218pt}
$
\left(\ddot{\bm{A}} - \nabla^2{\bm{A}}\right)\left( 1 + 2g^2|{\bf{H}}|^2\right) + 2g^2\partial_t(|{\bf{H}}|^2)\dot{\bm{A}}=0\,, \label{eq:F2} 
$
\end{adjustbox}
\end{equation}
where $|{\bf{H}}|^2 = H_{ij}H^{ij}$ is understood. In Sec. \,\ref{Sec.PR}, we found that parametric resonance occurs only for dark tensor homogeneous condensates possessing a null polarization state. From Eq.\,(\ref{Eq:H0}), we have $|{\bf{H}}|^2 = H^2(r)\text{cos}^2(\omega t)$. Thus,
\begin{align}
\left(\ddot{\bm{A}} - \nabla^2{\bm{A}}\right)\left( 1 + 2g^2H^2(r)\text{cos}^2(\omega t)\right) - 2\omega g^2H^2(r)\text{sin}(2\omega t)\dot{\bm{A}}=0\,, \label{eq:F2a}\\
\ddot{\bm{A}} - \nabla^2{\bm{A}} - 2\omega g^2H^2(r)\text{sin}(2\omega t)\dot{\bm{A}}=0\,,\label{eq:F2b}
\end{align}
where we have multiplied Eq.\,(\ref{eq:F2a}) by $(1 + 2g^2H^2(r)\text{cos}^2(\omega t))^{-1}$ and kept terms until $\mathcal{O}(g^2H^2(r))$ within the effective field theory framework.

The vector spherical harmonic decomposition of the vector potential reads as
\begin{equation}
    {\bm{A}}(t,{{\bm x}}) = \int_0^{\infty} \frac{dk}{2\pi}\sum_{l=1}^{\infty}\sum_{m=-l}^{l} \left[ v_{klm}(t){\bf{M}}_{lm}(k,{\bm x}) - w_{klm}(t){\bf{N}}_{lm}(k,{\bm x}) \right]\,,\label{eq:Axt}
\end{equation}
where  ${\bf{M}}_{lm}(k,\xvec)$ and ${\bf{N}}_{lm}(k,\xvec)$
are the vector spherical wavefunctions defined in terms of the vector spherical harmonics ${\bf{\Phi}}_{lm}({\bm{\hat{x}}}), {\bf{Y}}_{lm}({\bm{\hat{x}}}), {\bf{\Psi}}_{lm}({\bm{\hat{x}}})$  as 
\begin{align}
{\bf{M}}_{lm}(k,{\bm{x}}) &= -\frac{i j_l(kr)}{\sqrt{l(l+1)}} \,{\bf{\Phi}}_{lm}({\bm{\hat{x}}})\,,\\
{\bf{N}}_{lm}(k,{\bm{x}}) &= -\sqrt{l(l+1)}\frac{j_l(kr)}{kr} {\bf{Y}}_{lm}({\bm{\hat{x}}}) - \left( \sqrt{\frac{l+1}{l}}\frac{j_l(kr)}{kr} - \frac{j_{l+1}(kr)}{\sqrt{l(l+1)}}\right) {\bf{\Psi}}_{lm}({\bm{\hat{x}}})\,,
\end{align}
where $r=|{\bm{x}}|$ is the radial coordinate, $v_{lm}(k,t)$ and $w_{lm}(k,t)$ are the electromagnetic mode functions, and $j_l$ is the spherical Bessel function. The vector spherical wavefunctions hold the following properties:
$\nabla \times {\bf{M}}_{lm} = -ik {\bf{N}}_{lm}, \nabla \times {\bf{N}}_{lm} = ik {\bf{M}}_{lm}, \nabla \cdot {\bf{M}}_{lm} = \nabla \cdot {\bf{N}}_{lm} = 0$.

\begin{figure}
\centering
  \includegraphics[scale=0.37]{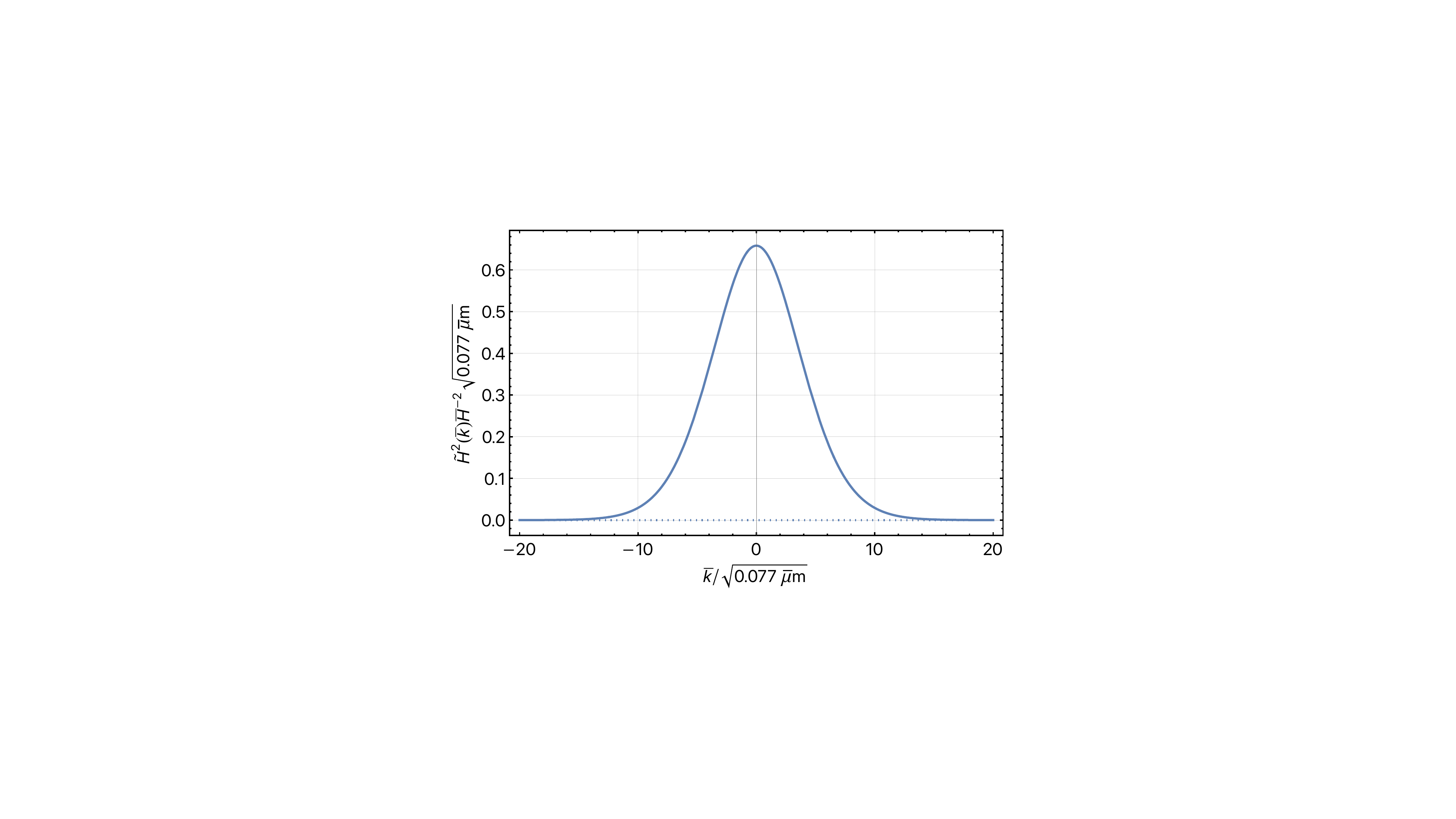}
\caption{\justifying The Fourier transform of the square of the tensor soliton profile $\widetilde{H^2}(\bar{k})\bar{H}^{-2}\sqrt{0.077\textcolor{black}{\bar\mu} m}$ in terms of the momentum $\bar{k}/\sqrt{0.077\textcolor{black}{\bar\mu} m}$. As the text mentions, the Fourier transform is peaked and centered at 0. This quantity plays a crucial  role for resonance in Eqs.\,(\ref{veqfinal}), (\ref{weqfinal}).}
 \label{FTHsquare}
\end{figure}

We replace Eq.~(\ref{eq:Axt}) into Eq.~(\ref{eq:F2b}). Since
${\bf{M}}_{lm}$ and ${\bf{N}}_{lm}$ satisfy the vector Helmoltz equation, the Laplacian of the vector potential follows the same equation, e.g.  $(\nabla^2 + k^2){\bm{A}} = 0$. Then,
\begin{align}
&\int_0^{\infty} \frac{dk}{2\pi}\sum_{l=1}^{\infty}\sum_{m=-l}^{l} \left[ 
\left( \ddot{v}_{klm}(t) + k^2{v}_{klm}(t) - 2\omega g^2H^2(r)\text{sin}(2\omega t)\dot{v}_{klm}(t) \right){\bf{M}}_{lm}(k,{\bm{\hat{x}}})  \right.\nonumber\\
&\textcolor{white}{xxxxxxxxxxxxxx}\left. -\left( \ddot{w}_{klm} + k^2{w}_{klm}(t) - 2\omega g^2H^2(r)\text{sin}(2\omega t)\dot{w}_{klm}(t) \right){\bf{N}}_{lm}(k,{\bm{\hat{x}}})\right] =0\,.
\end{align}
We use the well-known orthogonality properties of the vector spherical harmonics to select the equations for the electromagnetic modes labeled with $(l,m)$. To do that, we multiply via dot product the above equation with ${\bf{Y}}^{*}_{l'm'}$ and ${\bf{\Phi}}^*_{l'm'}$, and perform an angular integration. Recalling in particular that $\int {\bf{Y}}_{lm}\cdot{\bf{Y}}^{*}_{l'm'}d\Omega = \delta_{l l'}\delta_{m m'}$ and $\int {\bf{\Phi}}_{lm}\cdot{\bf{\Phi}}^*_{l'm'}d\Omega = l(l+1)\delta_{l l'}\delta_{m m'}$, we have 
\begin{align}
\int_0^{\infty} \frac{dk}{2\pi}  
\left( \ddot{v}_{klm}(t) + k^2{v}_{klm}(t) - 2\omega g^2H^2(r)\text{sin}(2\omega t)\dot{v}_{klm}(t) \right) \sqrt{l(l+1)} j_l(kr)& = 0 \,,\label{veq}\\
\int_0^{\infty} \frac{dk}{2\pi} \left( \ddot{w}_{klm}(t) + k^2{w}_{klm}(t) - 2\omega g^2H^2(r)\text{sin}(2\omega t)\dot{w}_{klm}(t) \right) \frac{\sqrt{l(l+1)}}{kr}j_l(kr) &=0\,.\label{weq}
\end{align}
We use the orthogonality of the spherical Bessel function to get rid of the integration over $k$ via a Dirac delta, e.g. 
$(2k^2/\pi)\int_0^{\infty}dr r^2 j_l(kr)j_l(k'r) = \delta (k-k')$, in the terms where the tensor soliton profile is absent. To do that, we multiply equations for the ${v}_{lm}(k,t)$ and ${w}_{lm}(k,t)$ modes by $j_l(k'r)r^2/\sqrt{l(l+1)}$ and $j_l(k'r)k'r^3/\sqrt{l(l+1)}$, respectively, and apply a radial integration. We obtain
\begin{align}
\ddot{v}_{klm}(t) + k^2v_{klm}(t) - \frac{4}{\pi}k^2\omega g^2\text{sin}(2 \omega t)
\int_0^{\infty} dk' \dot{v}_{k'lm}(t)  \int_0^{\infty} dr\, r^2 H^2(r) 
j_l(kr) j_l(k'r) &=0\,,\label{veq1}\\
\ddot{w}_{klm}(t) + k^2w_{klm}(t) - \frac{4}{\pi}k^3\omega g^2\text{sin}(2 \omega t)
\int_0^{\infty} \frac{dk'}{k'} \dot{w}_{k'lm}(t)  \int_0^{\infty} dr\, r^2 H^2(r) 
j_l(kr) j_l(k'r) &=0\,.\label{weq1}
\end{align}
Note that electromagnetic mode functions $w_{klm}(t)$ and $v_{klm}(t)$ do not couple between them so that Eqs.\,(\ref{veq1}, \ref{weq1}) may be treated independently. With the ultimate goal of taking care of the radial integral in the equations above, we express the square of the soliton radial profile in terms of its Fourier transform, e.g.  $H^2(r) = 2 \int_0^{\infty} d\bar k \widetilde{H^2}(\bar{k}) \text{cos}(\bar k r)/(2\pi)$, where we have extended the soliton profile domain by defining $H^2(-r)=H^2(r)$. Under this replacement, the radial integral is calculated as
\begin{align}
&\int_0^{\infty} dr r^2 H^2(r)j_l(kr)j_l(k'r) = \int_0^{\infty} dr r^2 j_l(kr)j_l(k'r) \int_0^{\infty} \frac{d\bar k}{\pi} \widetilde{H^2}(\bar k) \text{cos}(\bar k r)\,,\\
& =\frac{1}{2kk'}\int_{|k'-k|}^{k+k'} dk'' P_l \left( \frac{k^2+k'^2-k''^2}{2kk'} \right)
\int_{0}^{\infty} \frac{d\bar k}{\pi} \widetilde{H^2}(\bar k)
\int_0^{\infty} dr\, r \text{sin}(k'' r)\text{cos}(\bar k r)\,,\\
& =-\frac{\pi}{4kk'}\int_{|k'-k|}^{k+k'} dk'' P_l \left( \frac{k^2+k'^2-k''^2}{2kk'} \right) \frac{\partial}{\partial k''} 
\int_{0}^{\infty}  \frac{d\bar k}{\pi} \widetilde{H^2}(\bar k) \left[\delta(\bar k - k'') + \delta(\bar k + k'') \right]\,,\label{C17}\\
& =-\frac{1}{4kk'}\int_{|k'-k|}^{k+k'} dk'' P_l \left( \frac{k^2+k'^2-k''^2}{2kk'} \right) \frac{\partial}{\partial k''} 
 \widetilde{H^2}( k'') \,,
  \label{LegendreP}
\end{align}
where we have used the following identity for Bessel functions
\cite{Mehrem:2009ip} : 
\begin{equation}
j_l(kr)j_l(k'r) = \frac{1}{2kk'r} \int_{|k'-k|}^{k+k'} dk'' P_l \left( \frac{k^2+k'^2-k''^2}{2kk'} \right) \text{sin}(k'' r)\,.
\end{equation}
and $\int_0^{\infty} dr\, r  \text{sin}(k'' r) \text{cos}(\bar k r) = -(\pi/2) \frac{\partial}{\partial k''} \left[\delta(\bar k - k'') + \delta(\bar k + k'') \right]$. Note that in Eq.\, (\ref{C17}) the contribution only comes from $\delta(\bar k - k'')$ since $k''$ does not take negative values. Since we are dealing with configurations that slowly vary in space, the Fourier transform of their square $\widetilde{H^2}( k'')$ is centered at $k'' \sim 1/R$. From the resonance analysis in Sec. \ref{Sec.PR}, we infer that the resonance phenomenon occurs for wavenumbers $k \approx k' \approx \omega  \approx m$.
Since in the non-relativistic regime tensor solitons hold $m \gg 1/R$, the argument of the Legendre polynomial in Eq.\,(\ref{LegendreP}) is close to unity and $1-(k^2+k'^2-k''^2)/(2kk') \sim 1/(mR)^2$. Under this approximation and taking the expansion of the Legendre polynomial around one for a fixed $l$, we have 
\begin{align}
&\int_0^{\infty} dr r^2 H^2(r)j_l(kr)j_l(k'r)  \approx \nonumber \\
&\textcolor{white}{xxxx}-\frac{1}{4kk'}\int_{|k'-k|}^{k+k'} dk'' \left[  1 - \frac{l(l+1)}{(\sqrt{2}mR)^2} + \frac{l(l+1)(l+2)(l-1)}{(\sqrt{16}mR)^4} + \mathcal{O}\left(\frac{l}{mR}\right)^6  \right] \frac{\partial}{\partial k''} 
 \widetilde{H^2}( k'')\,.
\end{align}
In the regime at which $l \ll mR$, we can safely take the replacement $P_l \left( \frac{k^2+k'^2-k''^2}{2kk'} \right) 
\rightarrow 1$ to solve the radial integral from above as follows
\begin{align}
\int_0^{\infty} dr r^2 H^2(r)j_l(kr)j_l(k'r)  &\approx  -\frac{1}{4kk'}\int_{|k'-k|}^{k+k'} dk'' \frac{\partial}{\partial k''} \widetilde{H^2}( k'')\,,\nonumber \\
&\approx -\frac{1}{4kk'} \left( \widetilde{H^2}( k+k') - \widetilde{H^2}( k-k') \right) \approx 
\frac{1}{4kk'} \widetilde{H^2}( k-k')\,,
\end{align}
where we have neglected $\widetilde{H^2}( k+k')$, since $\widetilde{H^2}( \bar k)$ is centered at 0 as shown Fig.\, \ref{FTHsquare}. We have numerically checked that Eq.\,(\ref{LegendreP}) exponentially decays for large spherical harmonic number $l$ in the regime $l \gg mR$. Authors in Ref.\,\cite{Amin:2023imi} pointed out the same observation for the case of vector solitons.  Replacing the above equation in Eqs.\,(\ref{veq}, \ref{weq}), we finally have 
\begin{align}
\ddot{v}_{klm}(t) + k^2v_{klm}(t) - \omega g^2\text{sin}(2 \omega t)
\int_0^{\infty} \frac{dk'}{\pi} \dot{v}_{k'lm}(t) \widetilde{H^2}( k-k') 
 &=0\,,\label{veqfinal}\\
\ddot{w}_{klm}(t) + k^2w_{klm}(t) - \omega g^2\text{sin}(2 \omega t)
\int_0^{\infty} \frac{dk'}{\pi}  \dot{w}_{k'lm}(t)  \widetilde{H^2}( k-k') &=0\,,\label{weqfinal}
\end{align}
where we have applied the simplification $k \approx k' \approx m$. \textcolor{black}{The equations of motion for the electromagnetic mode functions share an identical form}, so that we just need to numerically solve Eq.\,(\ref{veqfinal}) or Eq.\,(\ref{weqfinal}). We numerically obtained the maximum real part of the Floquet exponent $\mu^{\text{(sol)}}_{\text{max}}$ via the discretization of the 1-dimensional $k$-space in a box and the perform of a generalized Floquet analysis, as explained in Sec.~(4.4) in~\cite{Hertzberg:2018zte}. We carefully set the width of the 1-dimensional box to be about $1/R$ centered at $k \approx m$ and sample it with a resolution of at least $O(g^2\bar{H}^2m/10)$. 

\begin{figure}
\centering
  \includegraphics[scale=0.32]{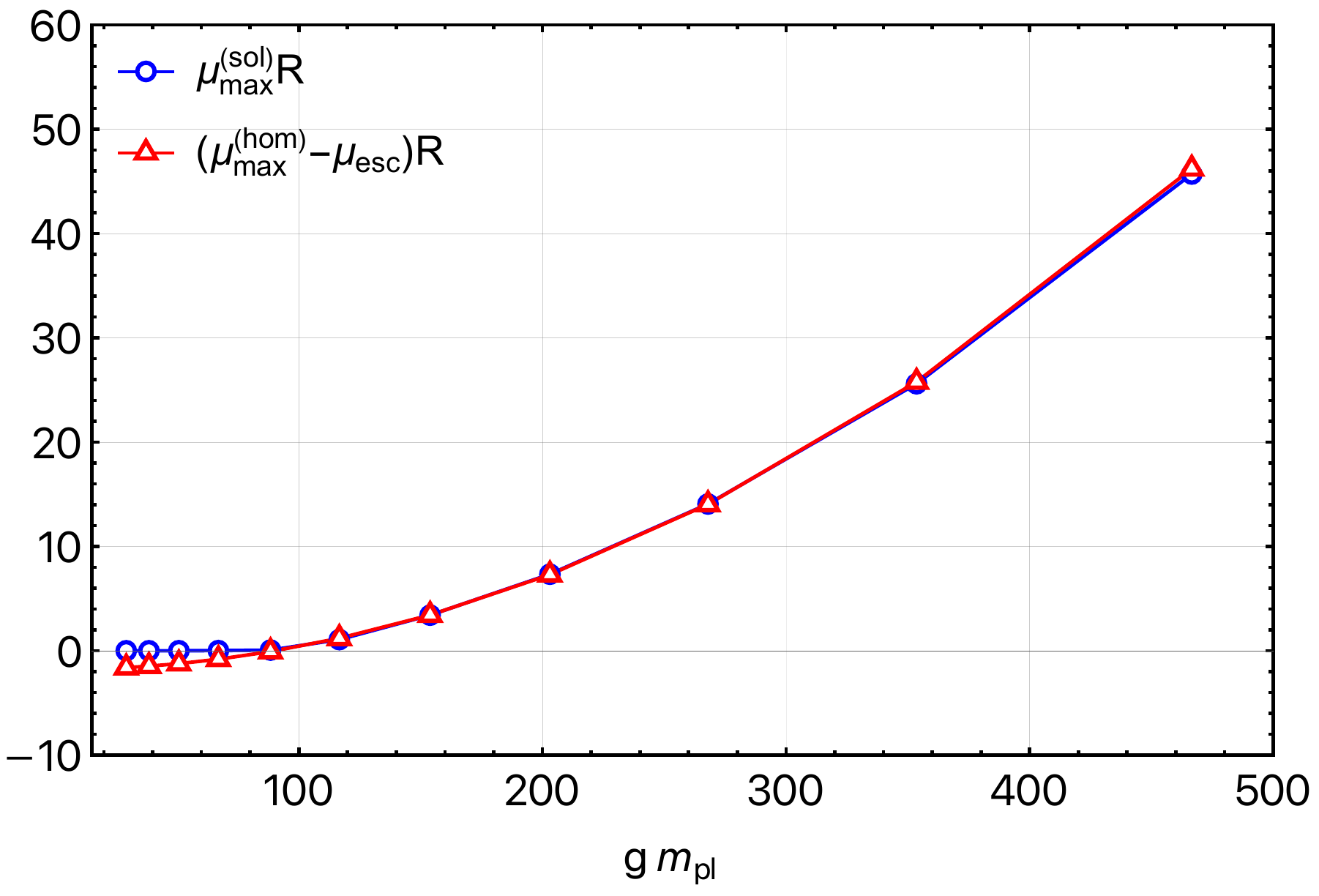}
\caption{\justifying  Numerical results from Eqs.\,(\ref{veqfinal}). Maximum real part of the Floquet exponent times the soliton radius $\mu^{\text{sol}}_{\text{max}} R_{\text{sol}}$ (blue points) for a tensor soliton with a number of particles $N_{\text{sol}}=m^2_{\text{pl}}/m^2$ and null polarization state. We assume that spin-2 particles couple to electromagnetism via the operator $\mathcal{O}_2 = -(1/2)F_{\mu\nu}F^{\mu\nu}H_{\alpha\beta}H^{\alpha\beta}$. Resonance is turned on when the maximum real part of the Floquet exponent for the corresponding homogeneous spin-2 condensate is comparable with the escape rate, $\mu_{\text{esc}}\approx 2/R_{\text{sol}}$, as shown by red points.}
 \label{ResSol}
\end{figure}

Figure \ref{ResSol} shows the numerical results (blue points) obtained by solving Eqs\,(\ref{veqfinal}) for the case of a tensor soliton with a number of particles 
$N_{\text{sol}}=m^2_{\text{pl}}/m^2$ and null polarization state. 
Before explaining its physical meaning, we give technical details for easy reproducibility.  The tensor soliton field amplitude, mass and radius are $\bar{H} = 2.1216 \times 10^{-3} m_{\text{pl}}$, $M_{\text{sol}} = 2.0\,  m^{2}_{\text{pl}}/m$ and $R_{\text{sol}} = 97.9874 m^{-1}$, respectively, for $\textcolor{black}{\bar\mu}/m = 1.04\times 10^{-3}$.  For each $g$ value, we define a 1-dimensional box in the $k$-space with 40 points in the interval ($k_{\text{min}},k_{\text{max}}$) around $k_{\text{res}}\approx m$. For example, for the point number 6th in the plot, we have $g = 116.634 m^{-1}_{\text{pl}}$, $1/R_{\text{sol}} = 0.0102054 m$, ($k_{\text{min}},k_{\text{max}}) = (0.92346,1.07654$), $k_{\text{max}}-k_{\text{min}}= 0.153081 m$, $\Delta k = 0.00392515$, and $(g \bar H)^2 = 0.0612323$. The $k$-space grid is able to cover the extension of $\widetilde{H}^{2}(k-k')$ and resolve the resonant band, e.g. 
$(k_{\text{max}}-k_{\text{min}}) > 1/R$ and $\Delta k < g^{2}\bar H^{2} m$.

Blue points in Fig. \ref{ResSol} tell us that, for a certain strength coupling, the resonance is turned on when the  
maximal real part of the Floquet exponent for the corresponding homogeneous spin-2 condensate, $\Hvec(t,\xvec) = \Hvec(t)$,  is comparable with the escape rate (red points), e.g. 
$\mu^{(\text{hom})}_{\text{max}} \approx \mu_{\text{esc}}$ with $ \mu_{\text{esc}}\approx 2/R_{\text{sol}}$.

\section{Galactic model and relative velocity distributions}
\label{App:GalacticModel}

We use the BCKM-model\,\cite{PhysRevD.87.083525} for the Milky Way Galaxy.
This model is constituted by a dark matter halo with a density profile  $\rho_{\text{DM}}(r)$ that follows a Navarro-French-White (NFW) profile\,\cite{1996ApJ...462..563N}, and a baryonic part formed by 
a spheroidal bulge with density $\rho_{\text{bulge}}(r)$ and exponential decay axisymmetric disk with density $\rho_{\text{disk}}(r)$ according to
\begin{align}
\rho_{\text{DM}}(r) &= \rho_{\text{DM},\,\odot} \left( \frac{R_0}{r} \right)\left( \frac{r_s + R_0}{r_s + r} \right)^2\,,\label{App:rhodm1}\\
\rho_{\text{bulge}}(r) &= \rho_{b,0} \left[ 1 + \left(\frac{r}{r_b}\right)^2  \right]^{-3/2}\,,\label{App:rhob}\\
\rho_{\text{disk}}(R,\,z)&= \frac{\Sigma_{\odot}}{2z_d}\exp\left(-\frac{R-R_0}{R_d}\right)\exp\left(-\frac{|z|}{z_d}\right)\,,
\label{App:rhodm2}
\end{align}
where $r \equiv |{\bf{x}}|$, $R_{\odot}$, $\Sigma_{\odot}$, $r_s$, $\rho_{b,0}$, $r_b$, $z_d$,  $R_d$, and $R=\sqrt{r^2-z^2}$ are the galactic radius, the solar distance from the center of the Galaxy,
the local surface density at the solar position,
the scale radius of the dark matter halo, the bulge's central density, the bulge's scale radius, the disk's scale height, the disk's scale radius, and the axisymmetric cylindrical coordinates, respectively.  By conducting a Markov Chain Monte Carlo analysis of observed Galactic rotation curve data, the authors in Ref.~\cite{PhysRevD.87.083525} reported the most likely values for the model parameters, which are presented in Table\,\ref{tab:gal-model}.
\begin{table}[t!]
\begin{tabular}{ |  m{2cm} |  m{2cm} | m{2cm}| m{2cm} | m{2cm} | m{2cm}| m{2cm} |} 
\toprule
\hline
\centering $r_{\text{vir}}$ &\centering $r_s$ & \centering $\rho_{\text{DM},\odot}$ & \centering $\rho_{b,0}$ & \centering $r_b$ & \centering $\Sigma_{\odot}$ &\textcolor{white}{xxxx} $R_d$ \\
 \centering $\left[\text{kpc}\right]$ &\centering $\left[\text{kpc}\right]$ &\centering $[\text{GeV/cm}^3]$ &\centering $[\text{GeV/cm}^3]$ &\centering   $\left[\text{kpc}\right]$ &\centering $[M_{\odot}/\text{pc}^2]$ & \textcolor{white}{ixxx}  $\left[\text{kpc}\right]$ \\
 \midrule
 \hline
  \hline
 \centering 199 &\centering 30.36 &\centering 0.19 &\centering $1.83\times10^{4}$ &\centering 0.092 &\centering 57.9 & \textcolor{white}{xxxxx}3.2 \\
 \bottomrule
 \hline
\end{tabular}
\caption{\justifying The most likely parameter values for the BCKM-model of the Milky Way obtained through a Markov Chain Monte Carlo
analysis using data from the Galactic rotation curve. These parameters are largely unaffected by the choice of $z_d$, which is fixed at 340\,\text{pc}\,\cite{PhysRevD.87.083525}.}
\label{tab:gal-model}
\end{table}

First, we calculate the velocity distribution functions associated with solitons in the Milky Way dark halo.  Using a spherically symmetric approximation\,\cite{Catena:2011kv}, we start computing the total galactic gravitational potential, $\Phi(r)$, which contains contributions from both dark (DM) and baryonic matter (BM) as follows.
\begin{equation}
\begin{split}
\Phi(r) &= \Phi_{\text{DM}}(r) +\Phi_{\text{BM}}(r)  
\approx \frac{1}{8\pi m^2_{\text{pl}}} \int_0^r \frac{M_{\text{DM}}(r') + M_{\text{BM}}(r')}{r'^2}\, \mathrm{d}r'\,,
\end{split}
\end{equation}
where the total mass of the dark and baryonic matter in the Galaxy is denoted by $M_{\text{DM}}$ and $M_{\text{BM}}$, respectively. It can be demonstrated that there is a one-to-one correspondence between the isotropic spatial density distribution \(\rho(r)\) and its corresponding phase-space distribution function \(\mathcal{F}(\xi)\) for any spherically symmetric system made up of collisionless components with isotropic velocity distribution functions, according to \cite{1916MNRAS..76..572E, 2008gady.book.....B}.
\begin{align}
\mathcal{F}(\xi) = \frac{1}{\sqrt{8}\pi^2} \left[ \int_0^{\xi} \frac{\mathrm{d}\Psi}{\sqrt{\xi-\Psi}}\frac{\mathrm{d}^2\rho}{\mathrm{d}\Psi^2} + \frac{1}{\sqrt{\xi}}\left(  \frac{\mathrm{d}\rho}{\mathrm{d}\Psi}\right)_{\Psi=0} \right]\,,\,\,\text{with}\,\,\,
\Psi(r) = -\Phi(r) + \Phi(r \rightarrow \infty)\,.
\label{eq:PSDF}
\end{align}
Here $\xi = \Psi(r) - v^2 / 2$ is the relative energy as a function of the relative potential $\Psi(r)$ and the velocity magnitude. Since we are interested in dark matter, we associate the dark matter halo profile at a certain radius $r$ with the soliton velocity distribution $f_r(\bm v)$ using the phase-space distribution function as $f_r(\bm v) = \mathcal{F}(\xi)/\rho_{\text{DM}}$, where 
\begin{equation}
  \int_0^\infty \int_\Omega f_r(\bm v) v^2 \, \mathrm{d} \Omega \, \mathrm{d} v = \int_0^\infty f_r(v)\,\mathrm{d} v = 1\,
\end{equation}
with $f_r(v)$ as the normalized speed distribution.

We compute the soliton velocity distribution functions and normalized speed distributions in the Galactic halo through Eq.\,(\ref{eq:PSDF}). Then we perform a series of Monte Carlo simulations by sampling $10^6$ velocities   
from the relevant velocity distributions, $f_r({\bm v})$. The relative speed distribution functions $f_r(v_{\text{rel}})$ are obtained by calculating the differences between these vectors using  Gaussian kernel density estimators.

\begin{figure}[h!]
\centering
  \includegraphics[scale=0.4]{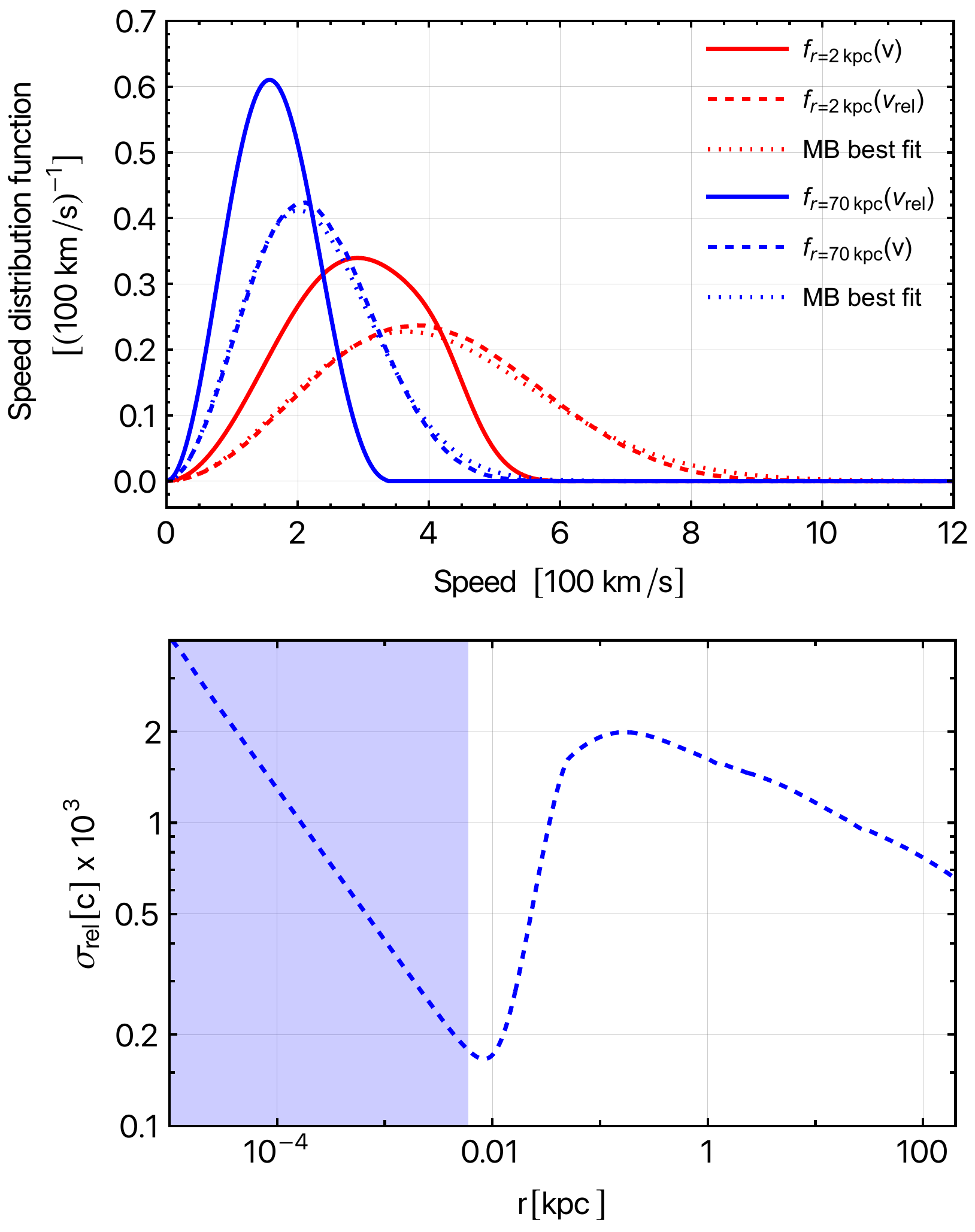}
\caption{\justifying (Top) Normalized relative speed distributions between solitons in the Galactic dark halo at two different Galactocentric radii (dashed lines) and their best-fits to the closest Maxwell-Boltzmann 3-d isotropic velocity distribution functions (pointed lines). The original normalized speed distributions for solitons are also shown (solid lines). (Bottom) The solitons' 3-d relative velocity dispersion in
light velocity units across the Galaxy. The colored area shows the 
inner region of the Galaxy, where the Jeans equation is used for calculation. We use the Eddington formalism for the outer regions, as the main text explains.}
 \label{Numericalfvrelsigmarel}
\end{figure}

Figure \ref{Numericalfvrelsigmarel} (top panel) shows the normalized relative speed distribution between solitons in the dark halo at Galactocentric radii $r = 2\,\text{kpc}$ (red dashed line) and 70 kpc (blue dashed line) and their corresponding best-fits to the closest Maxwell-Boltzmann 3-d isotropic velocity distribution functions (pointed blue and red lines). For completeness, we show the original normalized speed distributions for solitons at the studied radii. As the radius decreases, the typical relative speeds move toward larger values, and the distribution widens.   

In the inner regions of the Galaxy, we observe the presence of Sagittarius A*, a supermassive black hole located at the Galactic center. It has a mass of approximately \( M_{\text{SMBH}} \approx 3.5 \times 10^6 \, M_{\odot} \) and an associated gravitational influence radius \( r_h \) of about \( \sim 3 \, \text{pc} \) \cite{Safdi:2018oeu}. For sufficiently large distances (\( r \gg r_h \)), the influence of Sagittarius A* can be neglected; however, at distances where \( r \lesssim r_h \), its presence becomes significant for the calculation of relative velocities.
Due to the high computational cost, we migrate from the Eddington formalism to using the Jeans equation in regions very close to the Galactic center.  
We take the tensor soliton number density to follow a power-law holding an index $\bar{\gamma} = 1$ and use circular velocities $v_c(r) = ( M_{\text{gal}}(r)/(8\pi m^2_{\text{pl}}\, r))^{1/2}$, where $M_{\text{gal}}(r)$ is the total Galactic mass enclosed by a Galactic radius $r$. Assuming that velocities for solitons approximately follow Maxwell-Boltzmann 3-d isotropic velocity distribution functions, we have\,\cite{Dehnen:2006cm,Safdi:2018oeu}
\begin{equation}
\sigma \sim \frac{123\,\mathrm{km\,s^{-1}}}{\left(1+\bar{\gamma}\right)^{1/2}} \left( \frac{M_{\text{gal}}(r^*)}{r^*}\right)^{1/2}\,\,\,\,\longrightarrow\,\,\,\sigma_{\text{rel}} = \sqrt{2}\sigma\,,
\end{equation}
where $r^*=1\,\text{pc}$ and $\sigma$ ($\sigma_{\text{rel}}$) is the 3-d (relative) velocity dispersion for solitons.  

Figure \ref{Numericalfvrelsigmarel} (bottom panel) displays  $\sigma_{\text{rel}}$ in units of light velocity across the Milky Way. The colored area indicates the inner region of the Galaxy, where we use the Jeans equation for calculation. For the outer regions, we use the Eddington formalism. We observe that $\sigma_{\text{rel}}$ ranges approximately from $10^{-4}$ to $10^{-3}$ within the Galaxy. In the inner regions, the dispersion scales with the radius as $\sigma_{\text{rel}} \sim r^{-1/2}$, as the Galactic mass remains relatively constant due to the dominance of Sagittarius A*. 
\newpage

\bibliographystyle{JHEP.bst}
\bibliography{Tensor.bib}
\end{document}